\documentclass{article}
\usepackage{setspace}
\usepackage{arxiv}
\usepackage{xcolor}
\usepackage[utf8]{inputenc} 
\usepackage[T1]{fontenc}    
\usepackage{hyperref}       
\usepackage{url}            
\usepackage{booktabs}       
\usepackage{amsfonts}       
\usepackage{amsmath}
\usepackage{amssymb}
\usepackage{nicefrac}       
\usepackage{microtype}

\usepackage{tikz}
\usepackage{tikzscale}
\usetikzlibrary{math}
\usetikzlibrary{backgrounds}
\usepackage{pgfplots}
\pgfplotsset{compat = newest}
\usetikzlibrary{shapes.geometric}

\usepackage{verbatim}
\usepackage{nomencl}
\usepackage{subfig}
\usepackage{algorithm}
\usepackage[noend]{algpseudocode} 
\usepackage{rotating}
\usepackage{graphicx}
\usepackage{float}

\allowdisplaybreaks
\makenomenclature

\usepackage{color}

\title{Generative and discriminative training of Boltzmann machine through Quantum annealing}

\author{
  Siddhartha Srivastava\thanks{Corresponding author}\\
    Department of Mechanical Engineering\\
  University of Michigan Ann Arbor\\
   \texttt{sidsriva@umich.edu} \\
   \And
 Veera Sundararaghavan \\
  Department of Aerospace Engineering\\
  University of Michigan Ann Arbor\\
  \texttt{veeras@umich.edu} \\
}

\begin{document}
\maketitle

\begin{abstract}
A hybrid quantum-classical method for learning Boltzmann machines (BM) for a generative and discriminative task is presented. Boltzmann machines are undirected graphs with a network of visible and hidden nodes where the former is used as the reading site while the latter is used to manipulate visible states' probability. In Generative BM, the samples of visible data imitate the probability distribution of a given data set. In contrast, the visible sites of discriminative BM are treated as Input/Output (I/O) reading sites where the conditional probability of output state is optimized for a given set of input states. The cost function for learning BM is defined as a weighted sum of Kullback–Leibler (KL) divergence and Negative conditional Log-likelihood (NCLL), adjusted using a hyper-parameter. Here, the KL Divergence is the cost for generative learning, and NCLL is the cost for discriminative learning. A Stochastic Newton-Raphson optimization scheme is presented. The gradients and the Hessians are approximated using direct samples of BM obtained through Quantum annealing (QA). Quantum annealers are hardware representing the physics of the Ising model that operates on low but finite temperature. This temperature affects the probability distribution of the BM; however, its value is unknown. { \color{black}Previous efforts have focused on estimating this unknown temperature through regression of theoretical Boltzmann energies of sampled states with the probability of states sampled by the actual hardware. This assumes that the control parameter change does not affect the system temperature, however, this is not usually the case. Instead, an approach that works on the probability distribution of samples, instead of the energies, is proposed to estimate the optimal parameter set. This ensures that the optimal set can be obtained from a single run. A new technique to look for control parameters with better performance in training cost is also proposed where the KL divergence and NCLL are optimized with respect to the inverse system temperature and the result used to rescale the control parameter set. The performance of this approach as tested against the theoretically expected distributions show promising results for Boltzmann training on quantum annealers. }

\end{abstract}

\keywords{Quantum Machine learning \and Conditional Boltzmann Machines \and Annealing Temperature}

\section{Introduction}

Boltzmann machine (BM) is an energy-based model defined on an undirected graph and is used for unsupervised learning. The graph vertices are segregated into a set of visible and hidden nodes. The probability of each state is dependent on the total energy of the graph for that state. Moreover, only the state of a visible node is "visible" to the user. Therefore, these visible states' marginalized probabilities are a non-linear function of the energy parameters and can be used to model complicated distributions. These BMs can be trained either using Maximum-likelihood (ML) learning or Contrastive Divergence (CD) learning techniques. It is well known that ML Learning of Markov random fields (MRF) is a challenging task due to the large state space. Due to this complexity, Markov chain Monte Carlo (MCMC) methods typically take a long time to converge on unbiased estimates. CD learning, on the other hand, provides a computationally inexpensive way of training MRFs. However, it provides biased estimates in general \cite{carreira2005contrastive}.

\begin{figure}[tph]
\centering
\subfloat[General Boltzmann machine]{%
\includegraphics[width=0.25\linewidth]{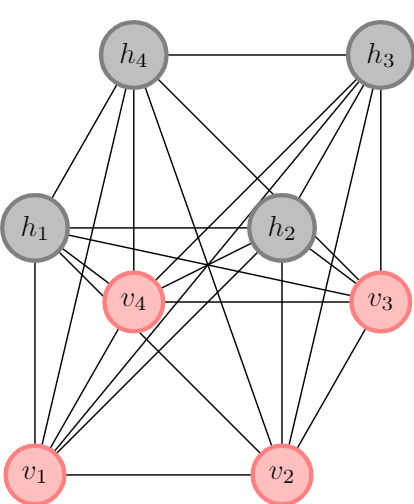}
}
\hspace{10mm}
\subfloat[Restricted Boltzmann machine]{%
\includegraphics[width=0.25\linewidth]{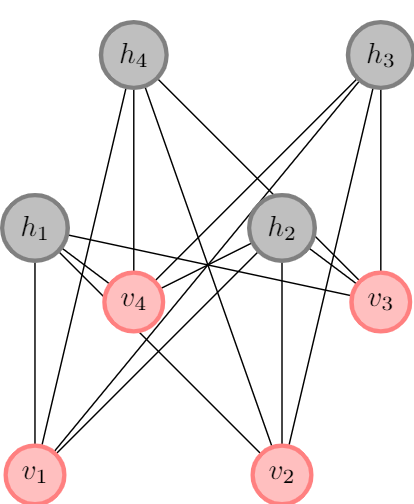}
}
\hspace{10mm}
\subfloat[Deep Boltzmann machine]{%
\includegraphics[width=0.25\linewidth]{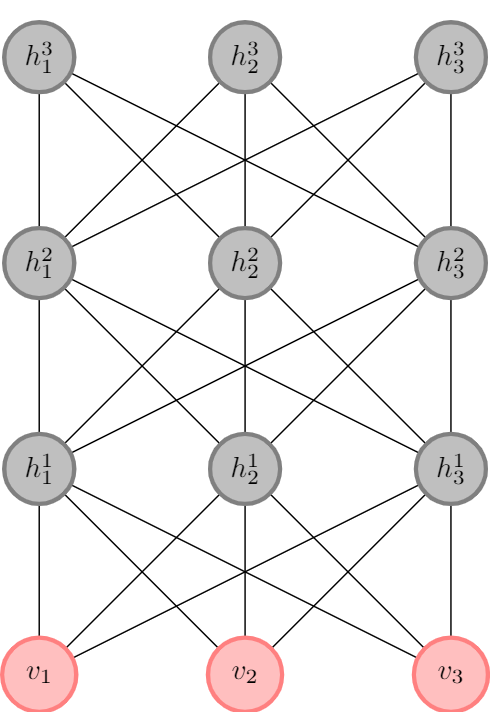}
}
\caption{\it Nomenclature of Boltzmann machines from \cite{salakhutdinov2009deep}}
\label{fig:nomenclature}
\end{figure}

A subclass of BM called the Restricted Boltzmann Machine (RBM) (see Fig\ref{fig:nomenclature}(b)) was proposed by Hinton (2002) \cite{hinton2002training} where the hidden and visible nodes had a bipartite structure. This structure allows an independent update of visible states, conditioned on the hidden states' knowledge and vice-versa. This property makes training of RBM very efficient on a classical computer. Boltzmann machines have received much attention as building blocks of multi-layer learning architectures for speech and image recognition \cite{jaitly2011learning, eslami2014shape}. The idea is that features from one RBM can serve as input to another RBM. By stacking RBMs in this way, one can construct the architecture of a Deep Boltzmann machine (see Fig\ref{fig:nomenclature}(c)). It is known that approximate inference in deep Boltzmann machines can handle uncertainty better and deal with ambiguous data \cite{salakhutdinov2009deep}.

A comparison between the ML and the CD-based training of RBM is presented in \cite{carreira2005contrastive}. The authors suggested that an initial CD-based training and a final ML-based fine-tuning of RBM is the most computationally efficient way of training RBMs with less bias. This bias issue was further studied in \cite{tieleman2008training} where the Persistent Contrastive Divergence (PCD) was developed. In this approach, the Markov Chain is not reset between parameter updates. This step brings the approximate gradient closer to the exact estimate in the limit of a small learning step. This method shows better performance on the testing data than the classical approach; however, it suffers from slow learning rates. A relatively faster approach was provided in
\cite{tieleman2009using} using the Fast Persistent Contrastive Divergence (FPCD) method. A tutorial on different training strategies is given in \cite{fischer2012introduction}. 

It is intuitive to see that General BM has more representative power than RBM and its derivatives. However, the efficiency of the above-mentioned training methods is not expected to translate to the general case as the data-dependent expectations are not easy to compute, at least using classical techniques. Quantum annealers (QA) have provided a promising way forward to tackle this problem of expectation estimation { \color{black}\cite{adachi2015application}}. QA are physical devices that operate on quantum mechanical laws and are designed to minimize the Ising model's energy { \color{black}\cite{kadowaki1998quantum}}. As they operate on finite temperatures, the simulations on QA results in sampling from the Boltzmann distribution of the Ising energies \cite{amin2015searching}. Researchers have recently employed this property of QA to train BMs with a slightly more complicated structure than RBMs. For instance, \cite{liu2020boltzmann} trained a Limited Boltzmann machine (LBM) to recover missing data in the images of chemical vapor deposition (CVD) growth for a $\text{MoS}_2$ monolayer. LBM allows sparse connectivity among the hidden units and, due to this complexity, it is not easy to deal with in a classical setting. 

Another direction that researchers have taken is the training of specialized RBMs that can be better represented on the QA architecture, e.g., the chimera RBM which emulates the topology of DWave Quantum annealers \cite{benedetti2016estimation}. This allows the model expectations to be estimated as a single sampling step instead of the k-step CD method. Meanwhile, the data-dependent expectations are estimated as the 1-step CD method due to the RBM's favorable topology. The result of this progress can be seen in the outburst of new applications of RBM in modern machine learning architectures, for instance, Sampling latent space in Quantum variational autoencoders (QVAE) \cite{khoshaman2018quantum}, RBM as an associative memory between generative and the discriminative part of the Associative Adversarial Network Model (AAN)  \cite{arici2016associative,wilson2019quantum} and Hybrid-Autoencoder-RBM approach to learn reduced dimension space \cite{sleeman2020hybrid}.

In this paper, an ML-based approach is studied for a General BM. As discussed earlier, the topology of a highly connected graph is not conducive for CD-based approaches. The major hurdle of generating independent samples of BM is circumvented using QA. At present, the two popular QA devices are the "DWave 2000Q" system with approximately 2000 qubits connected in a Chimera topology and the "DWave Advantage" system with approximately 5000 qubits connected in a Pegasus topology. Considering the physical devices' sparsity, the largest complete graph that can be simulated on these systems has size 64 on the 2000Q and 128 on the Advantage system. The past growth in these systems' computational power suggests the prospect of solving a large-scale problem in the near future. Taking the prospect for granted, large and arbitrarily connected BM can benefit from unbiased estimation via QA. The method developed in this work does not use the graph's topology and is numerically sub-optimal for the cases when such structures are present (e.g., bipartite graph). For such cases, the readers are encouraged to pursue the literature listed above and the bibliography therein. 

This paper aims to demonstrate the use of quantum annealers for discriminative and generative tasks involving Boltzmann machines. Generative tasks involve sampling a state from a probability distribution. At the same time, a discriminative BM acts as a classifier for a given dataset. A BM trained for generative and discriminative purposes can be used to sample a labeled dataset from a probability distribution. For example, \cite{dixit2020training} developed a generative BM for sampling vertical and horizontal stripe patterned images. { \color{black}The theoretical aspects of generative and discriminative training in the context of RBM can be found in \cite{larochelle2012learning}}. 
The second focus of this work is to analyze the effect of annealing temperature on training. The Boltzmann distribution is dependent on a sampling temperature, and the sampling temperature in QA is shown to be instance-dependent \cite{benedetti2016estimation}. { \color{black}An overview of few temperature-estimation methods for QA is provided in \cite{raymond2016global}. In a training strategy proposed by \cite{benedetti2017quantum}, the problem of temperature estimation is circumvented by assuming certain restrictions on the model and data distributions. However, their approach is machine-specific, in the sense that the knowledge of annealing temperature is required to use the trained model on a new QA machine.  
A temperature estimation strategy similar, in principle, to the one presented in \cite{korenkevych2016benchmarking} is adopted here.} However, an approach that works on the probability distribution of samples, instead of the energies, is proposed. This ensures that the optimal set can be obtained from a single run. A new technique to look for control parameters with better performance in training cost is also proposed where the KL divergence and NCLL are optimized with respect to the inverse system temperature. This method is employed to estimate the behavior of generative and discriminative costs and further refine the Ising model parameters.

\section{Notations and mathematical prerequisites} 

A Boltzmann Machine is a probabilistic graphical model defined on a complete graph which is partitioned into "visible" nodes taking up values observed during training denoted by the vector, $\boldsymbol{v}$, and "hidden" nodes where values must be inferred taking up values denoted by the vector, $\boldsymbol{h}$. These states collectively define the energy and consequently and the probability of each state. Next, the definition of a graph is stated to introduce useful terminology and notations. 

\subsection{Graph} 
A graph, $G$, is a pair of sets $(\mathcal{V}, \mathcal{C})$, where $\mathcal{V}$ is the set of vertices and $\mathcal{C}$ is the set of edges/connections. For each element $e\in \mathcal{C}$ there is a corresponding ordered pair $(x,y) ; x,y \in \mathcal{V}$ i.e. $\mathcal{C} \subseteq V\times V$. A Graph, $G=(\mathcal{V},\mathcal{C})$ is undirected if an edge does not have any directionality i.e $(x,y)\equiv (y,x)$. A graph is simple if $(x,x)\not\in \mathcal{C}$ for all $x\in \mathcal{V}$. The number of vertices are denoted by $N_V=|\mathcal{V}|$ and the number of edges are denoted by $N_C=|\mathcal{C}|$. The indices of connections and vertices are related using the maps, $\pi_1$ and $\pi_2$ such that for a connection with index, $k\in \{1,..,N_C\}$, the index of the corresponding vertices are $\pi_1(k)$ and $ \pi_2(k)$ with $1\leq \pi_1(k) < \pi_2(k) \leq N_V$. This essentially means $e_k \equiv (v_{\pi_1(k)}, v_{\pi_2(k)})$.

This work additionally requires the graph to be finite, i.e., $N_V<\infty$. Next, the definition of Ising energy is introduced. 

\subsection{Ising model}\label{sec:IsingModel}
Ising model is a type of a discrete pairwise energy model on an undirected simple graph, $G(\mathcal{V},\mathcal{C})$. Each vertex, $V_i\in V$ is assigned a state $s_i \in \lbrace 0,1\rbrace$ for all $i\in {1, \hdots ,N_V}$. This determines the complete state of the graph as an ordered tuple $\boldsymbol{S}=(s_1, \hdots ,s_i, \hdots , s_{N_V})\in \lbrace 0, 1 \rbrace^{N_V}$. The set of all possible states is referred to as $\mathcal{S} =\lbrace 0, 1 \rbrace^{N_V}$ with the total number of states denoted by $N_{TS} = |\mathcal{S}| = 2^{N_V}$. { \color{black}The Ising energy, $E$, for a particular state, $\boldsymbol{S}$ can be evaluated as follows:}
\begin{equation}\label{eq:IsingEnergy}
    E(\boldsymbol{S}) = \sum_{i=1}^{N_V} H_i  s_i + \sum_{k=1}^{N_C} J_{k} s_{\pi_1 (k)}  s_{\pi_2 (k)}
\end{equation}
where, the first terms represents the energy of labeling a vertex with label $s_i$, and the second term is the energy of labeling two connected vertices as $s_i$ and $s_j$. The parameters $H_i$ and $J_{k}$ are referred to as the Field strength and Interaction strength, respectively. 

The parameter set is represented as a vector, $\boldsymbol{\theta} =\begin{bmatrix}\theta_1, \hdots,\theta_{N_v + N_C}\end{bmatrix}^T$. In this work, it is specialized to following form: 
\begin{align*}
    \boldsymbol{\theta} = \begin{bmatrix}H_1, \hdots,H_{N_V},J_1,\hdots,J_{N_C}\end{bmatrix}^T
\end{align*}
This notation allows to describe energy as a matrix-product evaluated as $E(\boldsymbol{S}|\boldsymbol{\theta}) = \boldsymbol{\varepsilon}(\boldsymbol{S})\boldsymbol{\theta}$ where $\boldsymbol{\varepsilon}(\boldsymbol{S})$
\begin{align*}
    \boldsymbol{\varepsilon}(\boldsymbol{S}) = \begin{bmatrix} s_1, \hdots, s_{N_V},s_{\pi_1(1)}s_{\pi_2(1)},\hdots,s_{\pi_1(N_C)}s_{\pi_2(N_C)}\end{bmatrix}
\end{align*}

The distribution of equilibrated states can be modeled, at least approximately, as a Boltzmann distribution:
\begin{gather}\label{eq:boltzmann_prob}
    p(\boldsymbol{S}; \boldsymbol{\theta},\beta) = \frac{1}{Z} e^{-{E(\boldsymbol{S}|\boldsymbol{\theta})/k_B T}} \equiv \frac{1}{Z} e^{-\beta{E(\boldsymbol{S})}}
\end{gather}

Here, $Z$ denotes the partition function and is estimated as { \color{black}$Z = \sum_{\boldsymbol{S}} e^{-\beta{E(\boldsymbol{S})}} $}.

\subsection{Generative Boltzmann Machines}

The key idea behind a Boltzmann machine is the segregation of the vertices into visible and hidden states. This allows to write any state $\boldsymbol{S}$ of the graph as the following concatenation:
\begin{align*}
    \boldsymbol{S} = [\boldsymbol{v},\boldsymbol{h}]
\end{align*}
where $\boldsymbol{v}$ denotes the state of the visible nodes and $\boldsymbol{h}$ denotes the states of the hidden nodes. Only visible states are observed by the user and their probability can be estimated by marginalizing over all hidden states. Therefore, the probability of a particular visible state, $\boldsymbol{v}$, is given as,

\begin{equation}
p(\boldsymbol{v})= \sum_{\boldsymbol{h}} p(\boldsymbol{v}, \boldsymbol{h})= \frac{1}{Z} \sum_{\boldsymbol{h}} e^{-\beta E(\boldsymbol{v}, \boldsymbol{h})}
\end{equation}

This marginalization allows the BM to represent complex probability distributions. Consider a data set of $N_{DS}$ visible states, { \color{black}$\mathcal{D} = \{\boldsymbol{v}_{1}, \cdots, \boldsymbol{v}_{N_{DS}}\}$}. Each data state occurs with a probability $q(\boldsymbol{v}_{k})$ for all $k\in \{1,...,N_{DS}\}$, referred to as the true probability of the distribution. The performance of a BM can be judged by comparing the model distribution, $p(\boldsymbol{v})$ with the true distribution. This comparison can be carried out using the Kullback-Leibler divergence $D_{KL}(q||p)$ defined as,
\begin{gather*}
    D_{KL}(q||p; \boldsymbol{\theta},\beta) = -\sum_{\boldsymbol{v}\in \{\boldsymbol{v}_{1}, \cdots, \boldsymbol{v}_{N_{DS}}\}} q(\boldsymbol{v}) \ln{\frac{p(\boldsymbol{v}; \boldsymbol{\theta},\beta)}{q(\boldsymbol{v})}}
\end{gather*}
The KL divergence is always non-negative with $D_{KL}(q||p)=0$ if and only if $q=p$ almost everywhere. For this property, $D_{KL}$ is chosen to be the cost function for training generative BMs.

\textit{Remark}: In case, there is no meaningful notion of probability distribution of the data states. The true probability distribution can be substituted as $q(\boldsymbol{v}_{k}) = 1/N_{DS}$ for all $k\in \{1,\cdots,N_{DS} \}$. In this case, the KL Divergence is equal to the Log-likelihood of the data set normalized with the cardinality of the data set, $N_{DS}$. 

\subsection{Discriminative Boltzmann Machines}

It is often desired to generate a labelled data set which entails assigning a classification to each visible data point. This classifier can be included in our notation by further segregating the visible state into input-output pair. Consequently the state of the BM is represented as: 

\begin{align*}
    \boldsymbol{S} = [\boldsymbol{v}^I, \boldsymbol{v}^O, \boldsymbol{h}]
\end{align*}
where, $\boldsymbol{v}^I$ and $\boldsymbol{v}^O$ denotes the ``input'' and ``output'' visible state. The state, $\boldsymbol{v}^O$ is used to encode the classification of state $\boldsymbol{v}^I$. Discriminative BMs, also referred to as conditional BMs in literature, are trained for classification using labelled data set. The cost function in this case is taken as the Negative Conditional Log-likelihood $\mathcal{L}$ defined as,
\begin{gather*}
    \mathcal{N}(\boldsymbol{\theta},\beta) = -\sum_{[\boldsymbol{v}^I,\boldsymbol{v}^O]\in \{\boldsymbol{v}^{1}, ..., \boldsymbol{v}^{D}\}} \ln{p(\boldsymbol{v}^O|\boldsymbol{v}^I;\boldsymbol{\theta},\beta)}
\end{gather*}
where, the conditional probability, $p(\boldsymbol{v}^O|\boldsymbol{v}^I)$ is estimated as: 
\begin{align*}
    p(\boldsymbol{v}^O|\boldsymbol{v}^I) =\frac{p(\boldsymbol{v}^I,\boldsymbol{v}^O)}{\sum_{\widetilde{\boldsymbol{v}}^O} p(\boldsymbol{v}^I,\widetilde{\boldsymbol{v}}^O)} 
    \equiv \frac{\sum_{\boldsymbol{h}} p(\boldsymbol{v}^I,\boldsymbol{v}^O,\boldsymbol{h})}{\sum_{\widetilde{\boldsymbol{v}}^O,\widetilde{\boldsymbol{h}}} p(\boldsymbol{v}^I,\widetilde{\boldsymbol{v}}^O,\widetilde{h})} 
\end{align*}

\section{Training Method}

For a general purpose training strategy, the cost is set as a weighted average of KL Divergence and Negative conditional log-likelihood as described below 
\begin{align}\label{eq:Cost}
    \text{C} = \alpha D_{KL} + \frac{1-\alpha}{N_{DS}}\mathcal{N}(\boldsymbol{\theta}), \qquad \alpha \in [0,1]
\end{align}
where the $\alpha=0$ signifies a generative BM while $\alpha=1$ signifies a discriminative BM. Gradient based techniques are used to carry out the optimization procedure. The gradient is estimated as: 
\begin{align}\label{eq:Gradient}
    \frac{1}{\beta}\frac{\partial C}{\partial \theta_i} = -\alpha \mathbb{E}\left( \frac{\partial E}{\partial \theta_j}\right) +  \sum_{\boldsymbol{v}\in \mathcal{D}} \left( \left(\alpha q(\boldsymbol{v}) + \frac{1-\alpha}{N_{DS}}\right) \mathbb{E}\left( \left.\frac{\partial E}{\partial \theta_j}\right| \boldsymbol{v}\right) - \frac{1-\alpha}{N_{DS}}\mathbb{E}\left( \left.\frac{\partial E}{\partial \theta_j}\right|\boldsymbol{v^I}\right)\right) 
\end{align}
And, the hessian is estimated as: 
\begin{align}
    \frac{1}{\beta^2}\frac{\partial^2 C}{\partial \theta_i \partial \theta_j} = \alpha \operatorname{Cov}\left(\frac{\partial E}{\partial \theta_i},\frac{\partial E}{\partial \theta_j} \right)
    -\sum_{\boldsymbol{v}\in \mathcal{D}}
    \left(\alpha q(\boldsymbol{v}) + \frac{1-\alpha}{N_{DS}}\right) \operatorname{Cov}\left(\left.\frac{\partial E}{\partial \theta_i},\frac{\partial E}{\partial \theta_j}\right|\boldsymbol{v} \right) \nonumber\\
     +\sum_{\boldsymbol{v}\in \mathcal{D}}\frac{1-\alpha}{N_{DS}}\operatorname{Cov}\left(\left.\frac{\partial E}{\partial \theta_i},\frac{\partial E}{\partial \theta_j}\right|\boldsymbol{v}^I 
    \right) \label{eq:Hessian}
\end{align}
The definitions of all statistical quantities are presented in Appendix \ref{sec:stataQuantitites} and the derivatives of cost functions are estimated in Appendix \ref{sec:gradient_estimate}

\subsection{Optimization Scheme}

Stochastic gradient and Newton methods have been widely employed in such problems. A comparative performance of many variants of such stochastic methods are studied in \cite{loizou2020momentum}.  In stochastic optimization methods, it has been shown that both Newton methods and gradients-based methods have a local linear convergence rate. However, \cite{kovalev2019stochastic} developed a Newton method that is independent of the condition number in contrast to the gradient-based scheme. These developments motivate the use of Hessian in the optimization process. Such schemes are very useful in problems concerning sampling from sensitive devices like Quantum annealers. Analyzing the different variations of stochastic methods is out of scope of this work. A mini batch momentum-based approach is adopted. This approach can be easily substituted for one of the more sophisticated ones presented in \cite{loizou2020momentum,kovalev2019stochastic}. The following momentum-based update rule is used: 

\begin{equation}\label{eq:update_rule}
\boldsymbol{\theta}^{(t+1)}=\boldsymbol{\theta}^{(t)}+ \Delta \boldsymbol{\theta}^{(t)}, \qquad \Delta \boldsymbol{\theta}^{(t)} = \eta \boldsymbol{r}^{(t)}-\lambda \boldsymbol{\theta}^{(t)}+\nu \Delta \boldsymbol{\theta}^{(t-1)} 
\end{equation}

The parameter, $\eta$ defines the learning rate of the method and $\nu$ defines a momentum rate. A higher momentum rate means that the current learning direction is closer to the learning rate in the previous time step. In general, the momentum is kept low at the beginning and slowly increased as the learning progresses. This technique is employed to reduce oscillations in the final stages of training. The parameter $\lambda$ modifies the cost function to minimize the magnitude of the learned parameter. In this work. this parameter is identically set to $0$ in all test cases and is mentioned only for completeness. The variable, $\boldsymbol{r}$, denotes the rate of update. In the gradient-based method, it is estimated as:
\begin{align*}
    \boldsymbol{r}^{(t)} = -\nabla_{\boldsymbol{\theta}^{(t)}} C
\end{align*}
In Newton method, rate of update is estimated as: 
\begin{align*}
    \boldsymbol{r}^{(t)} = - (\nabla^2_{\boldsymbol{\theta}^{(t)}} C)^{-1} \nabla_{\boldsymbol{\theta}^{(t)}} C
\end{align*}

\textit{Remark 1}: The Hessian matrix estimated from the sampling process, is usually rank deficient. The main reason is under-sampling. Therefore, inversion of these matrices pose a challenge. In this work, Tikhonov regularization is used where singularity of $\nabla^2_{ \boldsymbol{\theta}^{(t)}} C$ is alleviated by adding positive terms to the diagonal as 
\begin{align*}
    \nabla^2_{\boldsymbol{\theta}^{(t)}} C \rightarrow \nabla^2_{\boldsymbol{\theta}^{(t)}} C + \epsilon^2 \mathbb{I}
\end{align*}

where $\mathbb{I}$ is the identity matrix. This regularization results in the following useful property for the rate of update: 
\begin{align*}
    \boldsymbol{r}^{(t)} = \operatorname{argmin}_{\widetilde{\boldsymbol{r}}} || (\nabla^2_{\boldsymbol{\theta}^{(t)}} C) \widetilde{\boldsymbol{r}} + \nabla_{\boldsymbol{\theta}^{(t)}} C||_2 + \epsilon^2||\widetilde{\boldsymbol{r}}||_2
\end{align*}

\textit{Remark 2}: The above update rule works for unconstrained optimization. A constrained optimization problem can be considered by employing lagrange multipliers. In this study, the constraints are much simpler, $|H_i|<H_{\text{0}}$ and $|J_k|<J_{\text{0}}$. These constraints represent the practical range of parameters for Quantum annealers. These bounds are implemented by using following scaling parameter:
\begin{align*}
 \delta = \max \left\lbrace \frac{\max_{i\in \{1,\hdots N_V\}} |H_i|}{H_0}, \frac{\max_{k\in \{1,\hdots N_C\}} |J_k|}{J_0} \right\rbrace 
\end{align*}
In any optimization step, if $\delta>1$, then the parameters are scaled as: $\boldsymbol{\theta}^{(t)} \rightarrow \boldsymbol{\theta}^{(t)}/\delta$ and the corresponding $\Delta\boldsymbol{\theta}^{(t)}$ is updated. The optimization procedure is presented in Algorithm \ref{Algo:optim}. The procedure uses a subroutine \textit{EstimateDerivatives} (Algorithm \ref{Algo:GradEstimation}) to estimate the gradients and hessian. This subroutine is discussed in the next section. For gradient based approaches, the estimation of hessian can be tactically avoided from algorithm \ref{Algo:GradEstimation} by ignoring all covariance terms and the hessian in this case is set to a Identity matrix. 

\begin{algorithm}[tph]
\caption{Optimization Step (per epoch)}\label{Algo:optim}
\begin{algorithmic}[1]
\State \textbf{Input}: $M$ (Number of Batches), $\boldsymbol{\theta}^{(t)}$ (Current Parameters), $\{\eta, \lambda, \nu\}$ (Learning parameters)
\State Partition $\mathcal{D}$ into $M$ partitions $\mathcal{D}_1, ..., \mathcal{D}_M$
\State For all $\{i\in 1,...,M\}$, define $q_i: \mathcal{D}_i \rightarrow \mathbb{R}$ such that $q_i(\boldsymbol{v}_k) = q(\boldsymbol{v_k})/\sum_{\boldsymbol{v}_j\in \mathcal{D}_i} q(\boldsymbol{v}_j)$ for all $\boldsymbol{v}_k \in \mathcal{D}_i$
\State $\boldsymbol{\theta}^{(t,0)} \gets \boldsymbol{\theta}^{(t-1)}$  
\State $\Delta\boldsymbol{\theta}^{(t,0)} \gets \Delta\boldsymbol{\theta}^{(t-1)}$  
\For {$i\gets 1 $ to $M$}
\State BatchData $\gets \{\mathcal{D}_i, q_i\}$
\State $\nabla_{\boldsymbol{\theta}^{(t,i)}} C, \nabla_{\boldsymbol{\theta}^{(t,i)}}^2 C \gets$ \textit{EstimateDerivatives}.BatchData \Comment{Algorithm \ref{Algo:GradEstimation}}
\State Estimate $\boldsymbol{r}^{(t,i)}$
\State $\Delta \boldsymbol{\theta}^{(t,i)} \gets \eta \boldsymbol{r}^{(t,i)} - \lambda \boldsymbol{\theta}^{(t,i)} + \nu \Delta \boldsymbol{\theta}^{(t,i-1)}$ 
\State $\boldsymbol{\theta}^{(t,i+1)} \gets \boldsymbol{\theta}^{(t,i)} + \Delta \boldsymbol{\theta}^{(t,i)}$
\State Apply constraints
\EndFor
\State $\boldsymbol{\theta}^{(t+1)} \gets \boldsymbol{\theta}^{(t,M+1)}$ 
\State $\Delta\boldsymbol{\theta}^{(t)} \gets \Delta \boldsymbol{\theta}^{(t,M+1)}$ 
\State \textbf{return} $\boldsymbol{\theta}^{(t+1)}$, $\Delta \boldsymbol{\theta}^{(t)}$
\end{algorithmic}
\end{algorithm}

\subsection{Numerical Estimation of Gradient and Hessian}

QA can be treated as a black-box that samples the states for a given set of parameters. DWave Quantum annealer is used in this work that operates based on the $\{+1,-1\}$ states. The energy parameters for the $\{+1,-1\}$ states and the \{1,0\} states can be transformed using the relation presented in Appendix \ref{sec:basis}\footnote{Basis transformation may scale the parameters outside the allowed range of DWave. This problem can be mitigated by choosing appropriate bounds on Field and Interaction parameters in the optimization process}. The user can specify the number of samples. The probability of a particular sample state is then be estimated as the ratio of the number of occurrences of that state to the specified number of samples. It is easily noticeable that the gradients and hessian described in terms of statistics of the energy gradient $\nabla_{\boldsymbol{\theta}} E$. The first term in the gradient and the hessian, requires estimation of $\mathbb{E}(\nabla_{\boldsymbol{\theta}} E)$ and $\operatorname{Cov}(\nabla_{\boldsymbol{\theta}} E)$. In the notation described in Section \ref{sec:IsingModel}, given a sample state $\boldsymbol{S}$, energy gradient is estimated as: 
\begin{align*}
    \nabla_{\boldsymbol{\theta}} E(\boldsymbol{S}) = \boldsymbol{\varepsilon}(\boldsymbol{S})
\end{align*}
The latter terms in Eq\eqref{eq:Gradient} and Eq\eqref{eq:Hessian} are more complicated to compute as they require summation over each visible data. Two possible strategies can be employed in this case. In the first approach, all sampled states are grouped according to the visible/input states. The conditional probabilities are then estimated by treating each data state's respective groups as the sample set. Since the samples from this estimation is unbiased, the QA samples are independent. In theory, the accuracy of these conditional probabilities increases with sample size. However, in practice, the number of samples is finite and should be kept to a minimum possible value to reduce computational cost. This is a critical drawback of this approach. For instance, not every data state needs to appear in the samples. In such cases, the KL Divergence cannot be estimated. 

\begin{figure}[tph]
\centering
\subfloat[Graph, $\mathcal{G}$]{%
\includegraphics[width=0.28\linewidth]{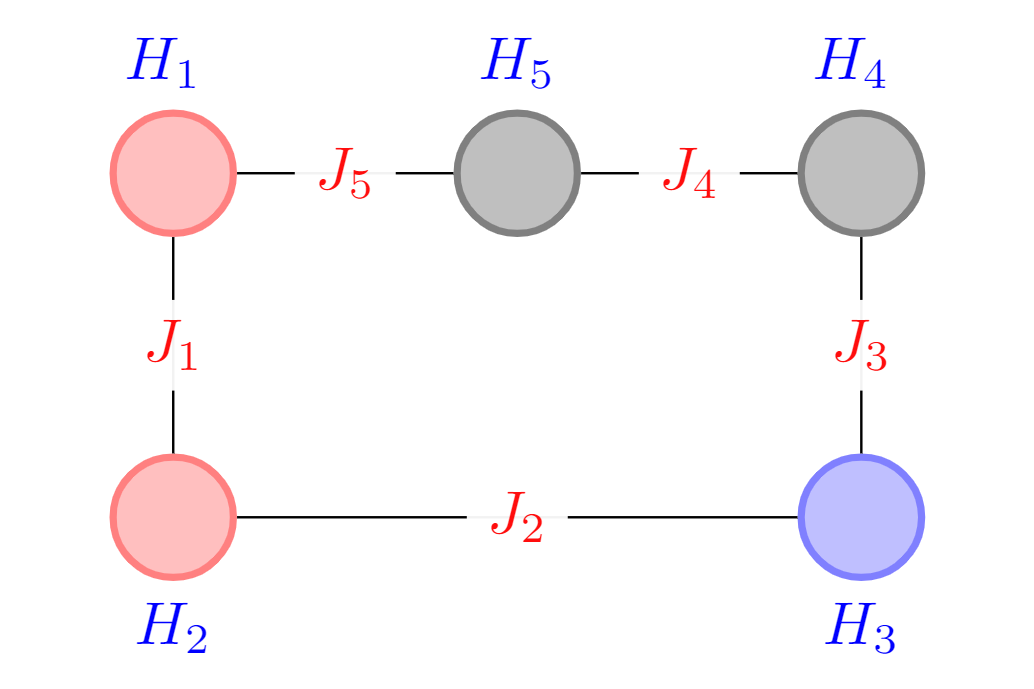}
}
\hspace{10mm}
\subfloat[Subgraph, $\mathcal{G}_{HO}$]{%
\includegraphics[width=0.28\linewidth]{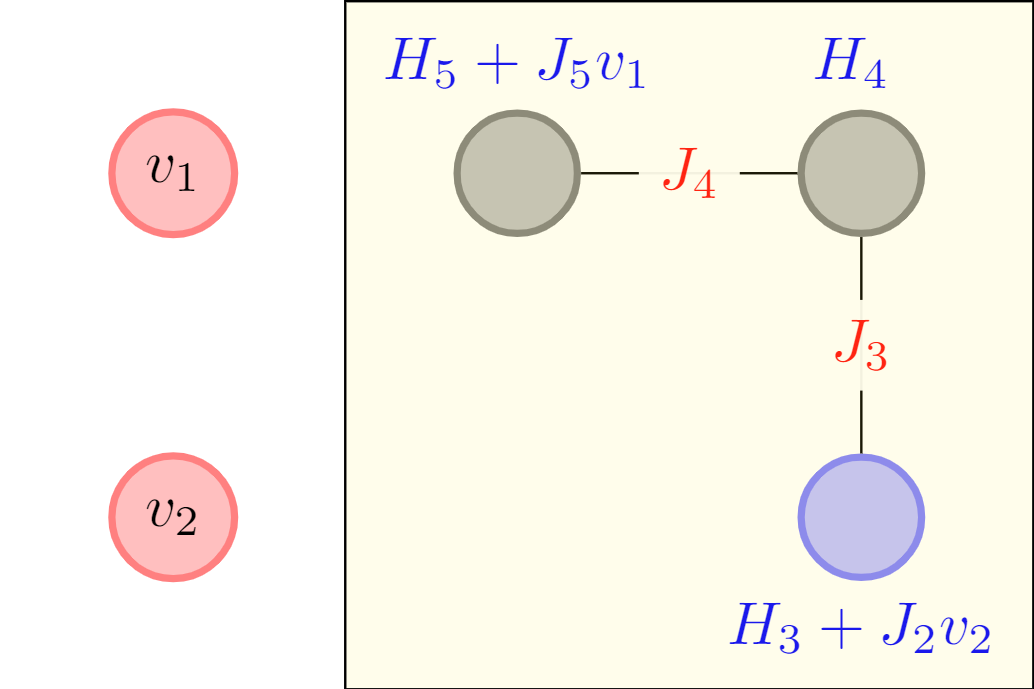}
}
\hspace{10mm}
\subfloat[Subgraph, $\mathcal{G}_H$]{%
\includegraphics[width=0.28\linewidth]{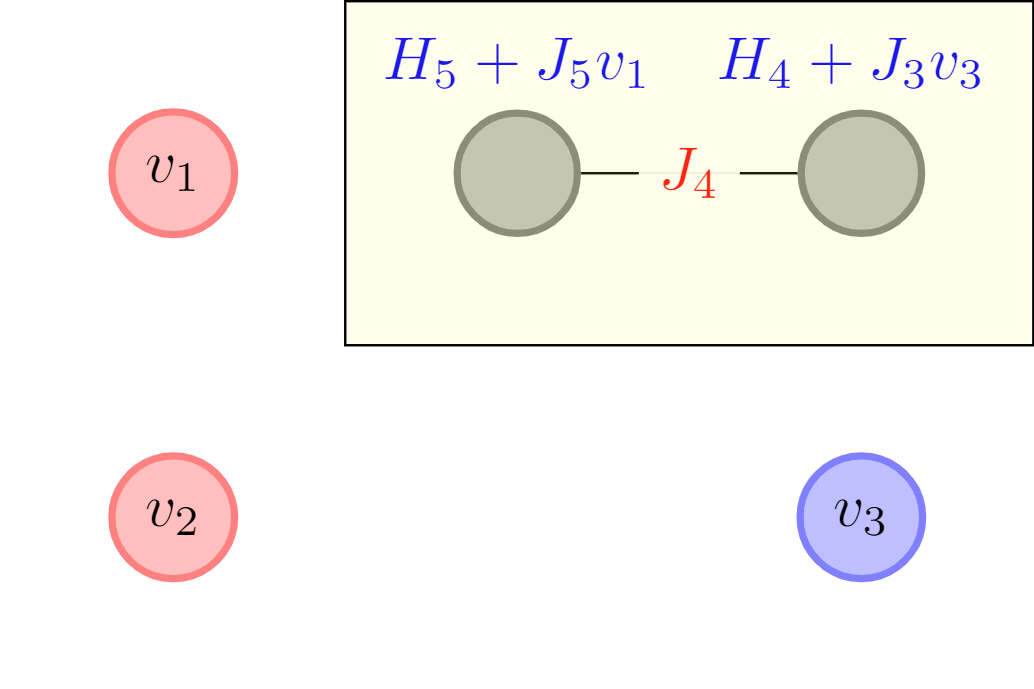}
}
\caption{Illustration for estimating parameters of the subgraph. The Input, Output and Hidden nodes are represented with red, blue and grey colors, respectively. The field parameters and interaction parameters are written in blue and red fonts, respectively. The subgraphs are presented in a yellow box. (a)  Cyclic graph, $\mathcal{G}$ with 2 input nodes, 1 output node and 2 hidden nodes (b) Subgraph of output and hidden nodes, $\mathcal{G}_{HO}$: Fixing the visible input $\boldsymbol{v}^I = [v_1, v_2]$ results in an augmented field term on the output and hidden nodes (c) Subgraph of hidden nodes, $\mathcal{G}_{H}$: Fixing the visible data $\boldsymbol{v} = [v_1, v_2, v_3]$ results in an augmented field term on the hidden nodes}
\label{fig:SubGraph}
\end{figure}

The other approach is to run independent simulations for each data state on the subgraph of hidden nodes, $\mathcal{G}_H$, and the subgraph of hidden and output nodes, $\mathcal{G}_{HO}$. The energy parameters of these subgraphs depend on the visible states (for $\mathcal{G}_H$), and input states (for $\mathcal{G}_{HO}$). The field terms are augmented to include the energy for fixing the states of the removed nodes. An illustration of this procedure is presented in Fig\ref{fig:SubGraph}. One can observe that this process leads to a shift in energy states. For instance, same states in Fig\ref{fig:SubGraph}(a)\&(b) have an energy difference of $H_1 v_1 + H_2v_2 + J_1 v_1 v_2$. However, the Boltzmann distribution remains unchanged by a uniform shift in energies. The drawback of this method is that this procedure's computational cost grows with the training data size. This growth by itself is not a problem; however, the sampling step is usually the most time-consuming. In general, CD-1 steps are used to determine this term in RBMs quickly. However, this method is not extendable to General BMs. The authors are currently unaware of any scheme that circumvents this computation. This second approach of running independent simulations for each data will be adopted from here onward. The procedure for estimating gradient and hessian from the sampled data is presented in Algorithm \ref{Algo:GradEstimation}. It should be noted that the estimation of RHS of Eq\eqref{eq:Gradient} and Eq\eqref{eq:Hessian} only yields the direction of gradient and hessian, respectively. The size of the update can be subsumed in the learning rates. However, the value of $\beta$ influences the probability distribution and, consequently, influences trained parameters' value. This issue is discussed in the next section. 

{ \color{black}It is also worth noting that the exact scaling of the computational complexity of these problems are very hard to ascertain, and most likely not well-defined. The reason being that the computational cost of estimation of gradient and hessian scales with the sample size while the optimal sample size is problem dependent \cite{ayanzadeh2020reinforcement}. However, for any graph with $N_V$ vertices and $N_S$ samples, the estimation of gradient requires $\mathcal{O}(N_S N_V)$ computations while the estimation of Hessian requires at most $\mathcal{O}(N_S N_V^2)$ computations. Due to limited number of vertices in quantum annealers ($\sim 10^3$) a quadratic cost is not expected to be a bottle neck in the near term.}

\begin{algorithm}[tph]
\caption{\textit{EstimateDerivative}: Estimation of Gradient and Hessian}\label{Algo:GradEstimation}
\begin{algorithmic}[1]
\State $\{\boldsymbol{v}_1, \hdots,\boldsymbol{v}_{DS}\}, \{q(\boldsymbol{v}_1), \hdots,q(\boldsymbol{v}_D)\}\gets$ Batch Data
\State $\{\boldsymbol{S}\} \gets$ Sample state of $\mathcal{G}$
\State Estimate random variable $\nabla_{\boldsymbol{\theta}}E$ from $\{\boldsymbol{S}\}$
\State Estimate $\mathbb{E}(\nabla_{\boldsymbol{\theta}}E)$, $\operatorname{Cov}(\nabla_{\boldsymbol{\theta}}E)$
\For {$i\gets 1 $ to $DS$}
  \State $[\boldsymbol{v}^I,\boldsymbol{v}^O]\gets \boldsymbol{v}_i$
  \State Update Parameters of $\mathcal{G}_{HO}$ and $\mathcal{G}_{H}$
  \State $\{\boldsymbol{h}\} \gets$ Sample state of $\mathcal{G}_H$
  \State Estimate random variable $\nabla_{\boldsymbol{\theta}}E (\boldsymbol{v})$ from $\{[\boldsymbol{v}^I,\boldsymbol{v}^O,\boldsymbol{h}]\}$
  \State Estimate $\mathbb{E}(\nabla_{\boldsymbol{\theta}}E|\boldsymbol{v})$, $\operatorname{Cov}(\nabla_{\boldsymbol{\theta}}E|\boldsymbol{v})$
  \State $\{[\widetilde{\boldsymbol{v}}^O,\widetilde{\boldsymbol{h}}]\} \gets$ Sample state of $\mathcal{G}_{HO}$
  \State Estimate random variable $\nabla_{\boldsymbol{\theta}}E (\boldsymbol{v}^I)$ from $\{[\boldsymbol{v}^I,\widetilde{\boldsymbol{v}}^O,\widetilde{\boldsymbol{h}}]\}$
  \State Estimate $\mathbb{E}(\nabla_{\boldsymbol{\theta}}E|\boldsymbol{v}^I)$, $\operatorname{Cov}(\nabla_{\boldsymbol{\theta}}E|\boldsymbol{v}^I)$
\EndFor
\State Evaluate Eq\eqref{eq:Gradient} and Eq\eqref{eq:Hessian}
\State \textbf{end}
\end{algorithmic}
\end{algorithm}

\subsection{Effect of Annealing Temperature}

Experimental evidence has shown that the apparent annealing temperature, i.e., the temperature corresponding to the Boltzmann distribution of samples, is instance-dependent \cite{benedetti2016estimation}. The corresponding inverse temperature is referred to as $\beta^*$ in this section. The consequence of this instance-dependence is that the quantum annealing systems cannot be rated for specific temperatures, and $\beta^*$ has to be estimated from the samples for each instance of the graph. The knowledge of $\beta^*$ is crucial in developing models capable of being implemented on different computational devices. Even in the same machine, two different embeddings of the same logical graph may lead to different annealing temperatures and consequently show disparities in performance. The key idea behind the estimation of $\beta^*$ is that the Boltzmann distribution of a state can be equivalently written as follows by taking a $\log$ on both sides:
\begin{align*}
    \log p(\boldsymbol{S}) = -\beta E(\boldsymbol{S}) - \log Z
\end{align*}
$\beta^*$ is estimated as the slope of this line. { \color{black}The exact form is as follows where $\mathbb{E}(.)$ denotes the expectation of the variable over all possible states: }
\begin{align}
    \beta^* = - \frac{\sum_{\boldsymbol{S}}  \left(E(\boldsymbol{S})-\mathbb{E}\left(E\right)\right) \left(\ln p(\boldsymbol{S}) - \mathbb{E}\left(\ln p\right) \right)  }{ \sum_{\boldsymbol{S}} (E(\boldsymbol{S})-\mathbb{E}(E))^2}
\end{align}
A similar approach was developed by \cite{benedetti2016estimation} that uses binning of energies. { \color{black}They estimated the value of $\beta$ using regression of ratio of probability of two bins of different energies that are sampled for two different control parameter sets.} The statistics of different energy levels can be succinctly written as: 
\begin{align*}
    p(E;\beta) = D_g(E) \frac{e^{-\beta E}}{Z}    
\end{align*}
where $D_g(E)$ is the degeneracy of states with energy, $E$. The authors used the log of ratio of probability, $l(\beta) = p(E_1;\beta)/p(E_2;\beta)$ for some fixed energy levels. They manipulated the value of $\beta$ by rescaling the parameters and were able to estimate the 
$\beta$ from the slope of $l(\beta)$ and the scaling factor. Readers should notice that although these two approaches are based on a similar argument, they differ greatly in their application. In the former approach, the intuition is that $\beta$ represents the slope of the following sampled data, $(E(\boldsymbol{S}),\log p(\boldsymbol{S}))$. Meanwhile, the latter approach uses the data of the probability distribution of energy levels. Both methods have their pros and cons. The second method requires sampling at rescaled parameters assuming that the effective annealing temperature remains invariant with rescaled parameters.  { \color{black} This is not usually the case as shown by \cite{raymond2016global} where a nonlinear relation was estimated between the rescaling of Energy and the effective $\beta$, attributed to `non-Boltzmann' nature of probability distribution. The variation in the distribution due to small changes in parameters is  overshadowed by the noise in the Boltzmann machine.  Meanwhile large parameter changes may lead to empty bins leading to an inability to compute probability ratios. For the first approach, that is proposed here, is equivalent to creating a bin of each unique state that is sampled, and this step is computationally more expensive than binning energy levels, especially in the limit of large graphs.  On the favorable side, it requires only one set of samples as a rescaling of parameters is not needed. Note that the probability of samples may be obtained as a direct output from current quantum annealing hardware, as is the case here in case of the D--wave annealer making this approach fairly easy to implement compared to the energy binning method.}

\section{Examples}\label{sec:Examples}

{ \color{black}The gradients for the cost as used in the training algorithm remain invariant to the system temperature upto a scaling as seen in Eq~ \eqref{eq:Gradient}. Assuming that distribution of states in the optimal solution can be effectively modelled as a Boltzmann distribution for some $\beta$, one can extract some useful statistics that can help further refine the model parameters for better performance (in terms of training cost). 

The key idea is that the variation of cost components ($D_{KL}$ and $\mathcal{N}$) with respect to $\beta$ (close to $\beta^*$) can be approximated using the statistics of samples obtained from the optimal parameter set. This information can elucidate in approximate sense, the $\beta$ at which the BM's performance is most optimal, say $\beta^o$, for a given choice of parameters ($\theta^*$). The user cannot control the annealing temperature but the parameters can rescaled to achieve the overall effect. It can be seen via Eq~\eqref{eq:boltzmann_prob} that scaling $\beta \rightarrow c\beta^*$ and $\boldsymbol{\theta}\rightarrow c\boldsymbol{\theta^*}$ has the same effect on the probability distribution. Since the hardware temperature cannot be modified, $\beta^o$ can then be used to rescale the parameters as $\boldsymbol{\theta}\rightarrow \boldsymbol{\theta} . \beta^o / \beta^*$.  to further reduce the cost. }

\subsection{Example: Temperature based scaling of training parameters}
\begin{figure}[tph]
\centering
\includegraphics[width=0.6\linewidth]{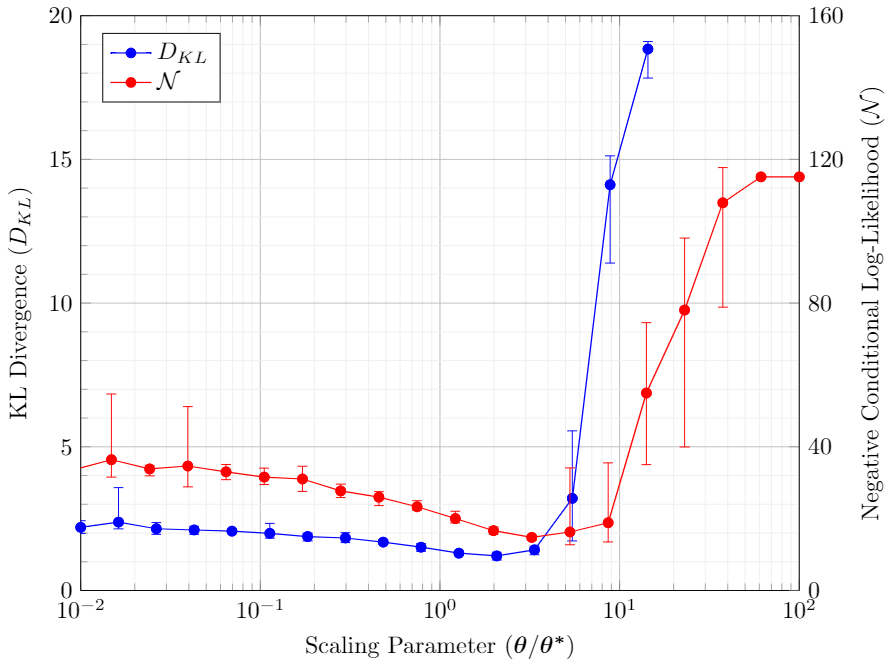}
\caption{Cost components for different scaling parameters. The initial set of parameters ($\boldsymbol{\theta^*}$) are for a BM with 10 vertices, trained for a 2-bit added circuit. The cost corresponds to the mean cost of 20 sample runs. The errorbar represents the minimum and maximum value in the sample set for each scaling parameter.
}
\label{fig:Scaling}
\end{figure}

{ \color{black}This effect is experimentally evaluated in Fig~\ref{fig:Scaling}, where the costs are computed for a fully connected graph with 10 vertices (7 visible and 3 hidden) trained with data representing the truth table of 2-bit adder circuit. The net cost of optimization is estimated for $\alpha=0.5$. Here the scaling factor of $1$ represents the initial parameter $\theta^*$. The individual cost components simultaneously decrease till the scaling factor of $~3$. These rescaled optimal parameters may not lie in the hardware-specified bounds, this may in fact be the reason why these parameters are not estimated by the training procedure in the first place. Next, we demonstrate a method which allows to evaluate this optimal scaling with only the statistics of samples at $\theta^*$. Unlike previous studies, this method assumes Boltzmann behavior only in the vicinity of $\beta^*$ while circumventing the problem of multiple evaluations of samples at different but close parameters. 
}

{ \color{black}Here, the Taylor expansions of KL Divergence and the Negative Conditional Log-likelihood as a function of  $\beta$ is employed to compute the optimal temperature. The advantage is that all the coefficients of  $\beta$ in the following expression can be estimated from the sampled states at $\beta^*$.} The Taylor expansion till the second term is as follows:  
\begin{align}\label{eq:taylor_approx}
    D_{KL}^{\text{app}} \left(\beta\right) &= D_{KL}^{*} + \left. \frac{\partial D_{KL}}{\partial \beta} \right|_{\beta^*} \left(\beta- \beta^* \right) + \frac{1}{2}\left. \frac{\partial^2 D_{KL}}{\partial \beta^2} \right|_{\beta^*} \left(\beta- \beta^* \right)^2 + \cdots \nonumber\\
    \mathcal{N}^{\text{app}} \left(\beta\right) &= \mathcal{N}^{*} + \left. \frac{\partial \mathcal{N}}{\partial \beta} \right|_{\beta^*} \left(\beta- \beta^* \right) + \frac{1}{2}\left. \frac{\partial^2 \mathcal{N}}{\partial \beta^2} \right|_{\beta^*} \left(\beta- \beta^* \right)^2 + \cdots 
\end{align}
where
\begin{align*}
    \frac{\partial D_{KL}}{\partial \beta} &= -\mathbb{E}_{\boldsymbol{v},\boldsymbol{h}}(E)+ \sum_{\boldsymbol{v}\in \{\boldsymbol{v}^{1}, ..., \boldsymbol{v}^{D}\}} q(\boldsymbol{v}) 
     \sum_{\boldsymbol{h}}  E(\boldsymbol{v},\boldsymbol{h})p(\boldsymbol{h}|\boldsymbol{v})\\
    \frac{\partial^2 D_{KL}}{\partial \beta^2} &= \sum_{\boldsymbol{v}\in \{\boldsymbol{v}^{1}, ..., \boldsymbol{v}^{D}\}} q(\boldsymbol{v}) \left( -\operatorname{Var}(E|\boldsymbol{v}) + \operatorname{Var}(E) \right)\\
    \frac{\partial \mathcal{N}}{\partial \beta} &= \sum_{[\boldsymbol{v}^I,\boldsymbol{v}^O]\in \{\boldsymbol{v}^{1}, ..., \boldsymbol{v}^{D}\}} \mathbb{E}(E|\boldsymbol{v}) - \mathbb{E}(E|\boldsymbol{v}^I) \\
    \frac{\partial^2 \mathcal{N}}{\partial \beta^2} &=  \sum_{[\boldsymbol{v}^I,\boldsymbol{v}^O]\in \{\boldsymbol{v}^{1}, ..., \boldsymbol{v}^{D}\}} -\operatorname{Var}(E|\boldsymbol{v}) + \operatorname{Var}(E|\boldsymbol{v}^I) 
\end{align*}

The exact behavior is estimated by evaluating the Boltzmann distribution, Eq\eqref{eq:boltzmann_prob}. The value of $\beta^o$ is estimated as the minimizer of the approximated quadratic cost function, $C^{\text{app}} = \alpha D_{KL}^{\text{app}} + (1-\alpha)\mathcal{N}^{\text{app}}/N_{DS} $. Therefore, when approximated cost is convex, the $\beta^o$ is estimated as:
\begin{align*}
    \beta^o = \beta^* -\left(  \frac{\partial^2 C}{\partial \beta^2} \right)^{-1} \frac{\partial C}{\partial \beta} 
\end{align*}

\begin{figure}[tph]
\centering
\includegraphics[width=0.6\linewidth]{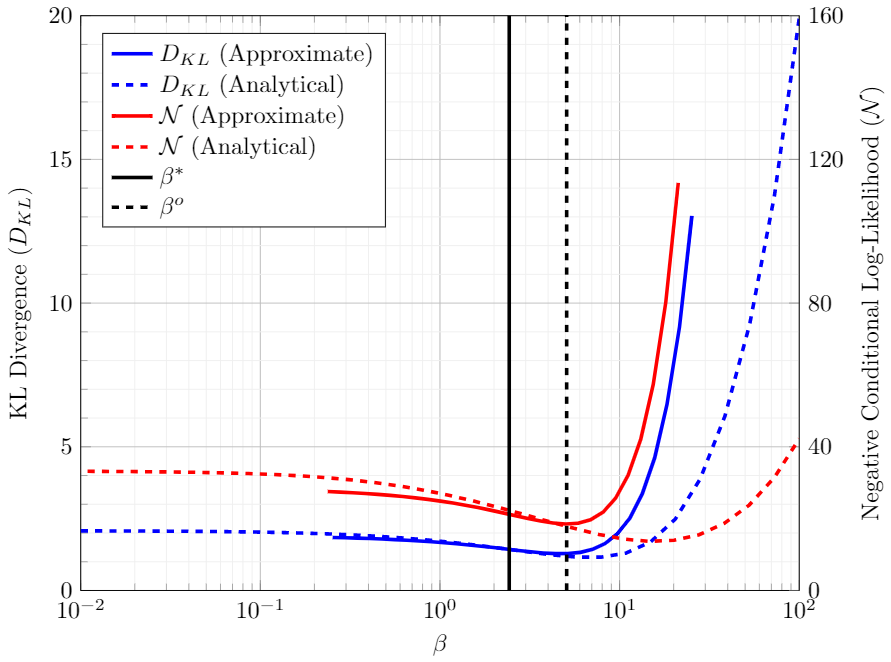}
\caption{Comparison of approximate and exact values of the training cost of BM with 10 vertices, trained for a 2bit added circuit.
}
\label{fig:LocalApproximation}
\end{figure}

{ \color{black}Figure~\ref{fig:LocalApproximation}, shows the individual cost components for the 2-bit adder example with respect to the value of $\beta$. The analytical model (dashed line) is estimated by enumerating all possible states of the Ising model and estimating the Boltzmann distribution. The approximated model is estimated using Eq~\eqref{eq:taylor_approx}. The estimated coefficients are: $\beta^* = 2.5251$, $\beta^o =5.2321 $, $D_{KL} = 1.4182$, $\mathcal{N} = 21.2122$, $\frac{\partial D_{KL}}{\partial \beta} = -0.1245$, $\frac{\partial^2 D_{KL}}{\partial \beta^2} = 0.0559$,  $\frac{\partial \mathcal{N}}{\partial \beta} = -1.9677$, $\frac{\partial^2 \mathcal{N}}{\partial \beta^2} = 0.7170$. The figure verifies the expected behaviour that both the components of the costs are simultaneously lowered at $\beta = \beta^o$. The cost components for rescaled parameters are $D_{KL} = 1.2193$, $\mathcal{N} = 16.5163$. The scaling factor for optimal parameters is approximated to be $\beta^*/\beta^o \approx 2.1$. This value is very close to the experimentally evaluated value of $\sim 3$. This procedure requires only $1$ set of samples in contrast to sampling at all possible $\beta$. } 

\subsection{Example: Generative and Discriminative training example}

\begin{figure}[tph]
\centering
\subfloat[Data points for generative training]{%
\includegraphics[width=0.28\linewidth]{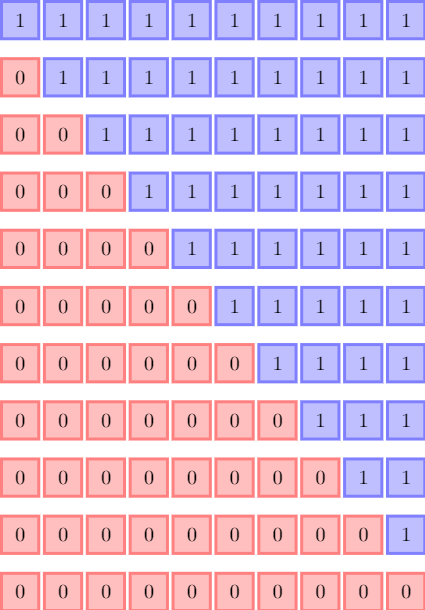}
}
\hspace{30mm}
\subfloat[Data points for discriminative training]{%
\includegraphics[width=0.28\linewidth]{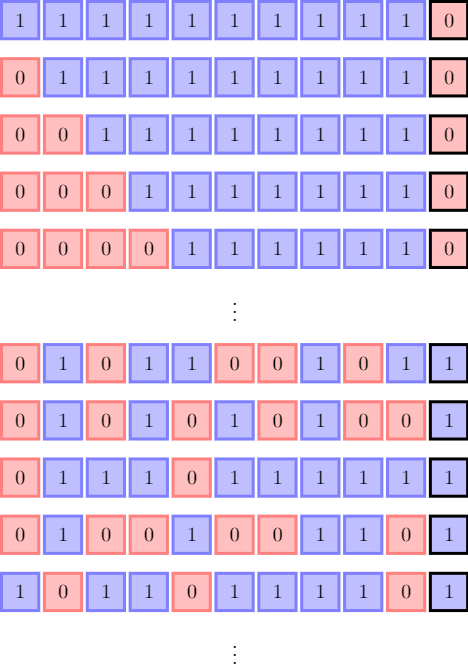}
}
\caption{Visible data states, $\mathcal{D}$, for training Boltzmann machines. Each row represents a single data point. (a) Each data point represents the phase of a state at 10 spatial points. The `0' phase is accumulated to the left, and the `1' phase is accumulated to the right with at most 1 boundary. (b) Labeled data set with the data points described in part(a) are labeled as '0', and data points with random spatial distribution labeled as `1'. The label is appended at the end of each data point in a black box. }
\label{fig:toyData}
\end{figure}

As a toy problem, the data illustrated in Fig\ref{fig:toyData} is used to train the BM models. The first data set (Fig\ref{fig:toyData}(a)) is unlabelled (Fig\ref{fig:toyData}(b)) and has 11 data points. The second data set has 40 data points with 11 labelled as `0' and 29 labelled as `1'. 

\begin{figure}[tph]
\centering
\includegraphics[width=0.6\linewidth]{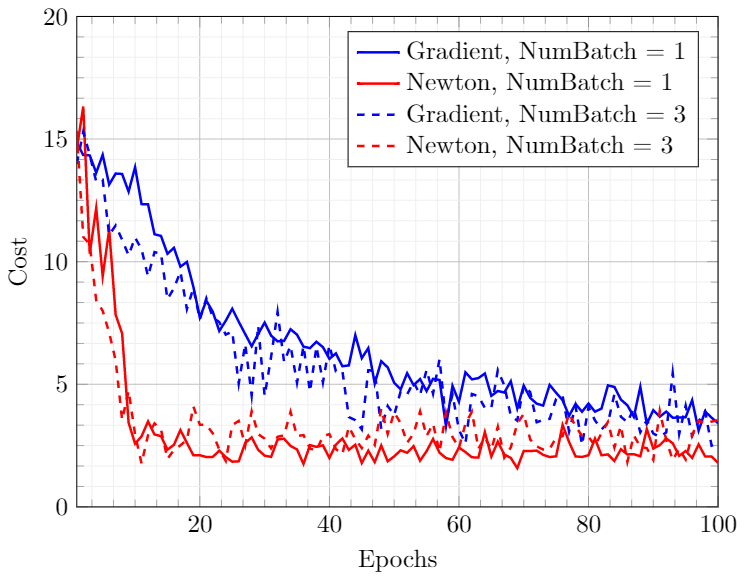}
\caption{Training data for a General BM with 3 Hidden nodes. The cost of training is defined by Eq\eqref{eq:Cost} with $\alpha = 0.5$.}
\label{fig:Training}
\end{figure}

In all the cases, the training parameters of Eq\eqref{eq:update_rule} have a constant value of $\eta = 0.1$, $\nu=0.7$, and $\lambda=0$. The Hessian is inverted using Tikhonov regularization with $\epsilon = 10^{-3}$. The run time data (see Fig \ref{fig:Training}) shows that the Newton approach performs better than the gradient-based approach. The fluctuations in the deterministic training case (NumBatch = 1) are due to the DWave sampling step's stochastic nature. It should also be remarked that the parameters were not optimized for individual cases and were arbitrarily picked from a suitable range. 

\begin{figure}[tph]
\centering
\includegraphics[width=0.6\linewidth]{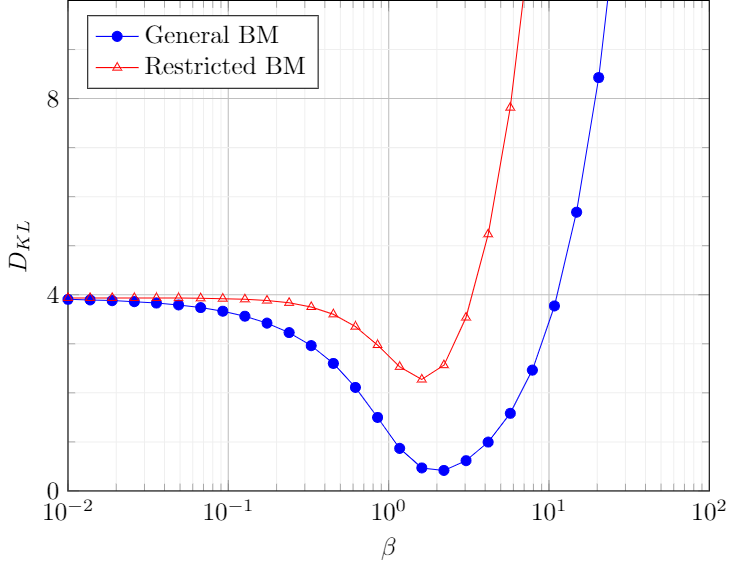}
\caption{Trained Generative BM with 3 hidden nodes}
\label{fig:GenBM}
\end{figure}

The stochastic training method with 2 batches was employed for the first data set (Generative learning). Two BM's with 3 hidden nodes were considered, first with complete connectivity and the second with Restricted BM architecture. 
The variation of KL Divergence (training Cost) with the annealing temperature is shown in Fig\ref{fig:GenBM}. It is observed that trained BM of the general type performs better than the Restricted type. This is an intuitive result as RBM is a specialized case of the General BM and has less representation capability in comparison.

The effect of cost parameter, $\alpha$ was studied for training General BM with the second dataset. The results are presented in Fig \ref{fig:GenBM}. The training is carried out with $70/30$ split into training and testing data. A large reduction in KL Divergence is observed, even for a small value of $\alpha$. Moreover, there was no substantial change in the conditional likelihood. This result suggests that the performance of Conditional/Discriminative BM can be enhanced by adding a small KL Divergence component to the training cost.

\begin{figure}[tph]
\centering
\includegraphics[width=0.6\linewidth]{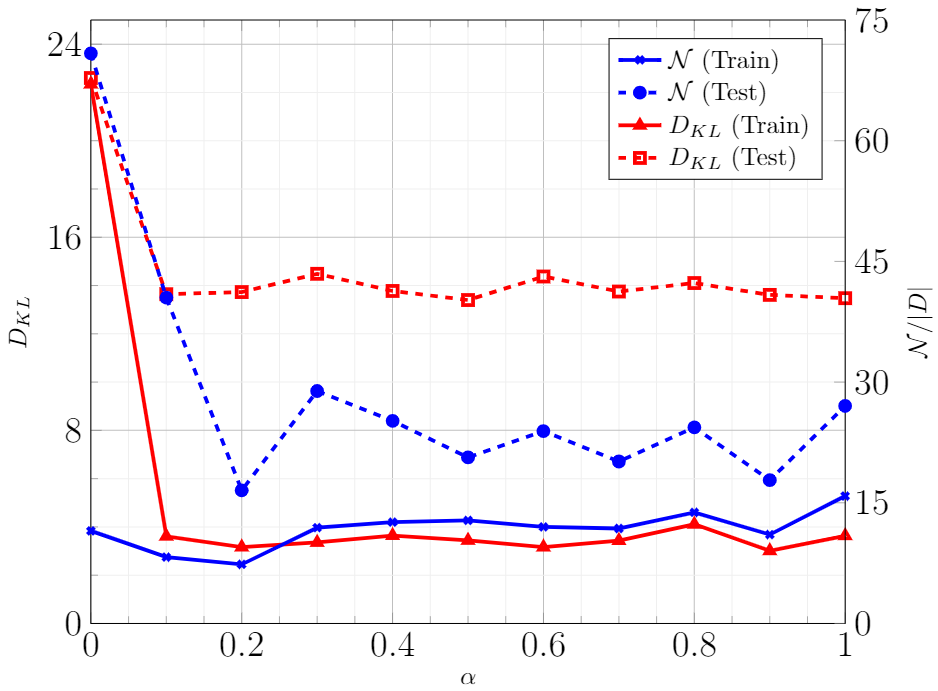}
\caption{Performance of trained BM with 4 hidden nodes with respect to the cost parameter, $\alpha$.}
\label{fig:GenDisBM}
\end{figure}

\section{Conclusion}

Quantum annealing has the potential to improve the training of General Boltzmann machines significantly. The stochastic Newton and gradient-based training methods can be employed using direct sampling from quantum states. { \color{black}In the proposed approach, inclusion of Hessian in training increases the computation cost from $\mathcal{O}(N_V)$ to $\mathcal{O}(N_V^2)$. For present Quantum annealers, this quadratic scaling does not pose any practical limitation due to limited number of qubits.} This procedure can accelerate the training of a General Boltzmann machine which have higher representation capability. 
The use of quantum annealers is promising for quantum/classical training since many qubits are available, and the training takes advantage of measurements on the physical realization of the Boltzmann machine \cite{khoshaman2018quantum,vinci2019path}. Unlike the other popular methods like the Contrastive Divergence, this method does not utilize the suggested BM's special topology. However, in practice, having a sparse connection is desirable as that allows embedding larger graphs in the DWave architecture. These methods were employed to carry out generative and discriminative training in toy problems. The numerical results suggested that stochastic Newton optimization performs better than gradient-based optimization. It was also observed that adding a small weightage to KL Divergence in discriminative cost greatly improves BM's performance. A major contribution of this paper is in developing a procedure to approximate the behavior of BM in slightly perturbed temperatures. Due to the instance-dependence of the QA sampler, even the same Ising problem with different hardware embedding may behave as two different quantum annealing system. { \color{black}Our procedure is useful in approximating a refined set of rescaled parameters for BM for a given embedding using the statistics of a single sample set of annealed states. In addition, once the hardware parameters are changed and optimized for a given generative or discriminative task using a gradient algorithm, the final parameter set can be further improved using a rescaling strategy. Here, the cost is additionally optimized with respect to the system temperature. It is shown that the approach allows improvement in Boltzmann machine training by comparing against exact analytical distributions. In the future, this work will be tested for other practically relevant benchmark problems alongwith rigorous analysis of training parameters as a function of hardware embedding. } 


\bibliographystyle{unsrt}  
\bibliography{references}  

\begin{thebibliography}{10}

\bibitem{carreira2005contrastive}
Miguel~A Carreira-Perpinan and Geoffrey~E Hinton.
\newblock On contrastive divergence learning.
\newblock In {\em Aistats}, volume~10, pages 33--40. Citeseer, 2005.

\bibitem{hinton2002training}
Geoffrey~E Hinton.
\newblock Training products of experts by minimizing contrastive divergence.
\newblock {\em Neural computation}, 14(8):1771--1800, 2002.

\bibitem{jaitly2011learning}
Navdeep Jaitly and Geoffrey Hinton.
\newblock Learning a better representation of speech soundwaves using
  restricted boltzmann machines.
\newblock In {\em 2011 IEEE International Conference on Acoustics, Speech and
  Signal Processing (ICASSP)}, pages 5884--5887. IEEE, 2011.

\bibitem{eslami2014shape}
SM~Ali Eslami, Nicolas Heess, Christopher~KI Williams, and John Winn.
\newblock The shape boltzmann machine: a strong model of object shape.
\newblock {\em International Journal of Computer Vision}, 107(2):155--176,
  2014.

\bibitem{salakhutdinov2009deep}
Ruslan Salakhutdinov and Geoffrey Hinton.
\newblock Deep boltzmann machines.
\newblock In {\em Artificial intelligence and statistics}, pages 448--455,
  2009.

\bibitem{tieleman2008training}
Tijmen Tieleman.
\newblock Training restricted boltzmann machines using approximations to the
  likelihood gradient.
\newblock In {\em Proceedings of the 25th international conference on Machine
  learning}, pages 1064--1071, 2008.

\bibitem{tieleman2009using}
Tijmen Tieleman and Geoffrey Hinton.
\newblock Using fast weights to improve persistent contrastive divergence.
\newblock In {\em Proceedings of the 26th Annual International Conference on
  Machine Learning}, pages 1033--1040, 2009.

\bibitem{fischer2012introduction}
Asja Fischer and Christian Igel.
\newblock An introduction to restricted boltzmann machines.
\newblock In {\em Iberoamerican Congress on Pattern Recognition}, pages 14--36.
  Springer, 2012.

\bibitem{adachi2015application}
Steven~H Adachi and Maxwell~P Henderson.
\newblock Application of quantum annealing to training of deep neural networks.
\newblock {\em arXiv preprint arXiv:1510.06356}, 2015.

\bibitem{kadowaki1998quantum}
Tadashi Kadowaki and Hidetoshi Nishimori.
\newblock Quantum annealing in the transverse ising model.
\newblock {\em Physical Review E}, 58(5):5355, 1998.

\bibitem{amin2015searching}
Mohammad~H Amin.
\newblock Searching for quantum speedup in quasistatic quantum annealers.
\newblock {\em Physical Review A}, 92(5):052323, 2015.

\bibitem{liu2020boltzmann}
Jeremy Liu, Ankith Mohan, Rajiv~K Kalia, Aiichiro Nakano, Ken-ichi Nomura,
  Priya Vashishta, and Ke-Thia Yao.
\newblock Boltzmann machine modeling of layered mos2 synthesis on a quantum
  annealer.
\newblock {\em Computational Materials Science}, 173:109429, 2020.

\bibitem{benedetti2016estimation}
Marcello Benedetti, John Realpe-G{\'o}mez, Rupak Biswas, and Alejandro
  Perdomo-Ortiz.
\newblock Estimation of effective temperatures in quantum annealers for
  sampling applications: A case study with possible applications in deep
  learning.
\newblock {\em Physical Review A}, 94(2):022308, 2016.

\bibitem{khoshaman2018quantum}
Amir Khoshaman, Walter Vinci, Brandon Denis, Evgeny Andriyash, Hossein Sadeghi,
  and Mohammad~H Amin.
\newblock Quantum variational autoencoder.
\newblock {\em Quantum Science and Technology}, 4(1):014001, 2018.

\bibitem{arici2016associative}
Tarik Arici and Asli Celikyilmaz.
\newblock Associative adversarial networks.
\newblock {\em arXiv preprint arXiv:1611.06953}, 2016.

\bibitem{wilson2019quantum}
Max Wilson, Thomas Vandal, Tad Hogg, and Eleanor Rieffel.
\newblock Quantum-assisted associative adversarial network: Applying quantum
  annealing in deep learning.
\newblock {\em arXiv preprint arXiv:1904.10573}, 2019.

\bibitem{sleeman2020hybrid}
Jennifer Sleeman, John Dorband, and Milton Halem.
\newblock A hybrid quantum enabled rbm advantage: convolutional autoencoders
  for quantum image compression and generative learning.
\newblock In {\em Quantum Information Science, Sensing, and Computation XII},
  volume 11391, page 113910B. International Society for Optics and Photonics,
  2020.

\bibitem{dixit2020training}
Vivek Dixit, Raja Selvarajan, Muhammad~A Alam, Travis~S Humble, and Sabre Kais.
\newblock Training and classification using a restricted boltzmann machine on
  the d-wave 2000q.
\newblock {\em arXiv preprint arXiv:2005.03247}, 2020.

\bibitem{larochelle2012learning}
Hugo Larochelle, Michael Mandel, Razvan Pascanu, and Yoshua Bengio.
\newblock Learning algorithms for the classification restricted boltzmann
  machine.
\newblock {\em The Journal of Machine Learning Research}, 13(1):643--669, 2012.

\bibitem{raymond2016global}
Jack Raymond, Sheir Yarkoni, and Evgeny Andriyash.
\newblock Global warming: Temperature estimation in annealers.
\newblock {\em Frontiers in ICT}, 3:23, 2016.

\bibitem{benedetti2017quantum}
Marcello Benedetti, John Realpe-G{\'o}mez, Rupak Biswas, and Alejandro
  Perdomo-Ortiz.
\newblock Quantum-assisted learning of hardware-embedded probabilistic
  graphical models.
\newblock {\em Physical Review X}, 7(4):041052, 2017.

\bibitem{korenkevych2016benchmarking}
Dmytro Korenkevych, Yanbo Xue, Zhengbing Bian, Fabian Chudak, William~G
  Macready, Jason Rolfe, and Evgeny Andriyash.
\newblock Benchmarking quantum hardware for training of fully visible boltzmann
  machines.
\newblock {\em arXiv preprint arXiv:1611.04528}, 2016.

\bibitem{loizou2020momentum}
Nicolas Loizou and Peter Richt{\'a}rik.
\newblock Momentum and stochastic momentum for stochastic gradient, newton,
  proximal point and subspace descent methods.
\newblock {\em Computational Optimization and Applications}, pages 1--58, 2020.

\bibitem{kovalev2019stochastic}
Dmitry Kovalev, Konstantin Mishchenko, and Peter Richt{\'a}rik.
\newblock Stochastic newton and cubic newton methods with simple local
  linear-quadratic rates.
\newblock {\em arXiv preprint arXiv:1912.01597}, 2019.

\bibitem{ayanzadeh2020reinforcement}
Ramin Ayanzadeh, Milton Halem, and Tim Finin.
\newblock Reinforcement quantum annealing: A hybrid quantum learning automata.
\newblock {\em Scientific reports}, 10(1):1--11, 2020.

\bibitem{vinci2019path}
Walter Winci, Lorenzo Buffoni, Hossein Sadeghi, Amir Khoshaman, Evgeny
  Andriyash, and Mohammad~H Amin.
\newblock A path towards quantum advantage in training deep generative models
  with quantum annealers.
\newblock {\em Machine Learning: Science and Technology}, 1(4):045028, 2020.

\end{thebibliography}


\appendix

\section{Definition of statistical quantities}\label{sec:stataQuantitites}
For completeness of notation, the definition of each statistical quantity is provided in context of the random variables used in this paper. Conditional quantities require following conditional probabilities: 
\begin{align*}
    &p(\boldsymbol{h}|\boldsymbol{v};\boldsymbol{\theta},\beta) = \frac{p([\boldsymbol{v},\boldsymbol{h}];\boldsymbol{\theta},\beta)}{\sum_{\widetilde{\boldsymbol{h}}}p([\boldsymbol{v},\widetilde{\boldsymbol{h}}];\boldsymbol{\theta},\beta)}\\
    &p([\boldsymbol{v}^O,\boldsymbol{h}]|\boldsymbol{v}^I;\boldsymbol{\theta},\beta) = \frac{p([\boldsymbol{v}^I,\boldsymbol{v}^O,\boldsymbol{h}];\boldsymbol{\theta},\beta)}{\sum_{\widetilde{\boldsymbol{v}}^O,\widetilde{\boldsymbol{h}}}p([\boldsymbol{v}^I,\widetilde{\boldsymbol{v}}^O,\widetilde{\boldsymbol{h}}];\boldsymbol{\theta},\beta)}
\end{align*}
\subsection{Expectations}
\begin{align*}
    &\mathbb{E}(E;\boldsymbol{\theta},\beta) = \sum_{\boldsymbol{S}} E(\boldsymbol{S};\boldsymbol{\theta}) p(\boldsymbol{S};\boldsymbol{\theta},\beta) \\
    &\mathbb{E}\left( \frac{\partial E}{\partial \theta_i};\boldsymbol{\theta},\beta\right) = \sum_{\boldsymbol{S}} \frac{\partial E}{\partial \theta_i}(\boldsymbol{S};\boldsymbol{\theta}) p(\boldsymbol{S};\boldsymbol{\theta},\beta) \\
    &\mathbb{E}\left( \left. \frac{\partial E}{\partial \theta_i} \right|\boldsymbol{v};\boldsymbol{\theta},\beta\right) = \sum_{\boldsymbol{h}} \frac{\partial E}{\partial \theta_i}([\boldsymbol{v},\boldsymbol{h}];\boldsymbol{\theta}) p(\boldsymbol{h}|\boldsymbol{v};\boldsymbol{\theta},\beta) \\
    &\mathbb{E}\left( \left. \frac{\partial E}{\partial \theta_i} \right|\boldsymbol{v}^I;\boldsymbol{\theta},\beta\right) = \sum_{\boldsymbol{v}^O,\boldsymbol{h}} \frac{\partial E}{\partial \theta_i}([\boldsymbol{v}^I,\boldsymbol{v}^O,\boldsymbol{h}];\boldsymbol{\theta}) p([\boldsymbol{v}^O,\boldsymbol{h}]|\boldsymbol{v}^I;\boldsymbol{\theta},\beta) 
\end{align*}
\subsection{Covariances}
The dependence on $\boldsymbol{\theta}$ and $\beta$ is dropped for notational convenience. 
\begin{align*}
    &\operatorname{Cov}\left(\frac{\partial E}{\partial \theta_i},\frac{\partial E}{\partial \theta_j} \right) = \sum_{\boldsymbol{S}} \frac{\partial E}{\partial \theta_i} \frac{\partial E}{\partial \theta_j}(\boldsymbol{S}) p(\boldsymbol{S}) -\mathbb{E}\left( \frac{\partial E}{\partial \theta_i}\right) \mathbb{E}\left( \frac{\partial E}{\partial \theta_j}\right)\\
    &\operatorname{Cov}\left(\left.\frac{\partial E}{\partial \theta_i},\frac{\partial E}{\partial \theta_j}\right|\boldsymbol{v} \right) = \sum_{\boldsymbol{h}} \frac{\partial E}{\partial \theta_i}\frac{\partial E}{\partial \theta_j}([\boldsymbol{v},\boldsymbol{h}]) p(\boldsymbol{h}|\boldsymbol{v}) - \mathbb{E}\left( \left. \frac{\partial E}{\partial \theta_i} \right|\boldsymbol{v}\right)\mathbb{E}\left( \left. \frac{\partial E}{\partial \theta_j} \right|\boldsymbol{v}\right)\\
    &\operatorname{Cov}\left(\left.\frac{\partial E}{\partial \theta_i},\frac{\partial E}{\partial \theta_j}\right|\boldsymbol{v}^I \right)= \sum_{[\boldsymbol{v}^O,\boldsymbol{h}]} \frac{\partial E}{\partial \theta_i}\frac{\partial E}{\partial \theta_j}([\boldsymbol{v}^I,\boldsymbol{v}^O,\boldsymbol{h}]) p([\boldsymbol{v}^O,\boldsymbol{h}]|\boldsymbol{v}^I) - \mathbb{E}\left( \left. \frac{\partial E}{\partial \theta_i} \right|\boldsymbol{v}^I\right)\mathbb{E}\left( \left. \frac{\partial E}{\partial \theta_j} \right|\boldsymbol{v}^I\right)\\
\end{align*}

\subsection{Variances}
\begin{align*}
    &\operatorname{Var}\left( E \right) = \sum_{\boldsymbol{S}} E^2(\boldsymbol{S}) p(\boldsymbol{S}) - \mathbb{E}^2\left( E\right)\\
    &\operatorname{Var}\left(\left.E\right|\boldsymbol{v} \right) = \sum_{\boldsymbol{h}} E^2([\boldsymbol{v},\boldsymbol{h}]) p(\boldsymbol{h}|\boldsymbol{v}) - \mathbb{E}^2\left( \left. E \right|\boldsymbol{v}\right)\\
    &\operatorname{Var}\left(\left.E\right|\boldsymbol{v}^I \right)= \sum_{[\boldsymbol{v}^O,\boldsymbol{h}]} E^2([\boldsymbol{v}^I,\boldsymbol{v}^O,\boldsymbol{h}]) p([\boldsymbol{v}^O,\boldsymbol{h}]|\boldsymbol{v}^I) - \mathbb{E}^2\left( \left. E\right|\boldsymbol{v}^I\right)\\
\end{align*}

\section{Estimation of gradients}\label{sec:gradient_estimate}

\subsection{Gradient of KL Divergence}
The gradient of Log-likelihood for a single data is estimated as:
\begin{equation}
\begin{array}{l}
{
\begin{aligned} \frac{\partial \ln p(\boldsymbol{v} )}{\partial \theta_j}
&=\frac{\partial}{\partial \theta_j}\left(\ln \sum_{\boldsymbol{h}} e^{-\beta E(\boldsymbol{v}, \boldsymbol{h})}\right)
-\frac{\partial}{\partial \theta_j}\left(\ln \sum_{\boldsymbol{v}', \boldsymbol{h}'} e^{-\beta E(\boldsymbol{v}', \boldsymbol{h}')}\right) \\
&=- \beta\sum_{\boldsymbol{h}} \frac{ e^{-\beta E(\boldsymbol{v}, \boldsymbol{h})}}{\sum_{\boldsymbol{h}'} e^{-\beta E(\boldsymbol{v}, \boldsymbol{h}')}} \frac{\partial E(\boldsymbol{v}, \boldsymbol{h})}{\partial \theta_j}
+\beta \sum_{\boldsymbol{v}', \boldsymbol{h}'}  \frac{ e^{-\beta E(\boldsymbol{v}', \boldsymbol{h}')}}{\sum_{\boldsymbol{v}'', \boldsymbol{h}''} e^{-\beta E(\boldsymbol{v}'', \boldsymbol{h}'')}} \frac{\partial E(\boldsymbol{v}', \boldsymbol{h}')}{\partial \theta_j} \\
&=- \beta\sum_{\boldsymbol{h}} \frac{\frac{1}{Z} e^{-\beta E(\boldsymbol{v}, \boldsymbol{h})}}{\frac{1}{Z}\sum_{\boldsymbol{h}'} e^{-\beta E(\boldsymbol{v}, \boldsymbol{h}')}} \frac{\partial E(\boldsymbol{v}, \boldsymbol{h})}{\partial \theta_j}
+\beta \sum_{\boldsymbol{v}', \boldsymbol{h}'}  \frac{ e^{-\beta E(\boldsymbol{v}', \boldsymbol{h}')}}{Z} \frac{\partial E(\boldsymbol{v}', \boldsymbol{h}')}{\partial \theta_j} \\
&=-\beta \sum_{\boldsymbol{h}} \frac{p(\boldsymbol{v} , \boldsymbol{h})}{p( \boldsymbol{v})} \frac{\partial E(\boldsymbol{v}, \boldsymbol{h})}{\partial \theta_j}+\beta \sum_{\boldsymbol{v}', \boldsymbol{h}'} p(\boldsymbol{v}', \boldsymbol{h}') \frac{\partial E(\boldsymbol{v}', \boldsymbol{h}')}{\partial \theta_j}\\
&=-\beta \sum_{\boldsymbol{h}} p(\boldsymbol{h} | \boldsymbol{v}) \frac{\partial E(\boldsymbol{v}, \boldsymbol{h})}{\partial \theta_j}+\beta \sum_{\boldsymbol{v}', \boldsymbol{h}'} p(\boldsymbol{v}', \boldsymbol{h}') \frac{\partial E(\boldsymbol{v}', \boldsymbol{h}')}{\partial \theta_j} \\
&=\beta\left( \mathbb{E}\left( \frac{\partial E}{\partial \theta_j}\right) - \mathbb{E}\left( \left.\frac{\partial E}{\partial \theta_j}\right|\boldsymbol{v}\right)\right) 
\end{aligned}}
\end{array}
\end{equation}
The gradient of KL Divergence is estimated as: 
\begin{equation}
\begin{array}{l}
{
\begin{aligned} 
    \frac{\partial D_{KL}(q||p)}{\partial \theta_j} &= -\sum_{\boldsymbol{v}\in \{\boldsymbol{v}^{1}, ..., \boldsymbol{v}^{D}\}} q(\boldsymbol{v}) \frac{\partial }{\partial \theta_i}\left( \ln{\frac{p(\boldsymbol{v})}{q(\boldsymbol{v})}} \right)\\
    &= \beta \left(-\mathbb{E}\left( \frac{\partial E}{\partial \theta_j}\right)+\sum_{\boldsymbol{v}\in \{\boldsymbol{v}^{1}, ..., \boldsymbol{v}^{D}\}} q(\boldsymbol{v}) \mathbb{E}\left( \left.\frac{\partial E}{\partial \theta_j}\right|\boldsymbol{v}\right) \right)
\end{aligned}}
\end{array}
\end{equation}

\subsection{Gradient of Negative Conditional Log-likelihood}

\begin{equation}
\begin{array}{l}
{
\begin{aligned} \frac{\partial\mathcal{N}}{\partial \theta_j}
&= - \sum_{[\boldsymbol{v}^I,\boldsymbol{v}^O]\in \{\boldsymbol{v}^{1}, ..., \boldsymbol{v}^{D}\}}
\left(  \frac{\partial \ln \sum_{\boldsymbol{h''}} e^{-\beta E(\boldsymbol{v}^I,\boldsymbol{v}^O,\boldsymbol{h}')} }{\partial \theta_j} -  \frac{\partial \ln \sum_{\boldsymbol{v'}^O,\boldsymbol{h}'} e^{-\beta E(\boldsymbol{v}^I,\boldsymbol{v'}^O,\boldsymbol{h'})} }{\partial \theta_j}\right)\\

&= \beta \sum_{[\boldsymbol{v}^I,\boldsymbol{v}^O]\in \{\boldsymbol{v}^{1}, ..., \boldsymbol{v}^{D}\}}\left( \sum_{\boldsymbol{h''}} \frac{Zp_\theta(\boldsymbol{v}^I,\boldsymbol{v}^O,\boldsymbol{h''})}{Zp_\theta(\boldsymbol{v}^I,\boldsymbol{v}^O)}\frac{\partial E(\boldsymbol{v}^I,\boldsymbol{v}^O,\boldsymbol{h}'')} {\partial \theta_j} - \sum_{\boldsymbol{v'}^O,\boldsymbol{h}'} \frac{Zp_\theta(\boldsymbol{v}^I,\boldsymbol{v'}^O,\boldsymbol{h}')}{Zp_\theta(\boldsymbol{v}^I)}\frac{\partial E(\boldsymbol{v}^I,\boldsymbol{v'}^O,\boldsymbol{h}')} {\partial \theta_j}
\right)\\

&=\beta \sum_{[\boldsymbol{v}^I,\boldsymbol{v}^O]\in \{\boldsymbol{v}^{1}, ..., \boldsymbol{v}^{D}\}} \left(
\sum_{\boldsymbol{h}''} p(\boldsymbol{h}''|\boldsymbol{v}^I,\boldsymbol{v}^O) \frac{\partial E(\boldsymbol{v}^I,\boldsymbol{v}^O, \boldsymbol{h}'')}{\partial \theta_j} - 
\sum_{\boldsymbol{v'}^O,\boldsymbol{h}'} 
p(\boldsymbol{v'}^O ,\boldsymbol{h}' | \boldsymbol{v}^I) \frac{\partial E(\boldsymbol{v}^I,\boldsymbol{v'}^O, \boldsymbol{h}')}{\partial \theta_j} 
\right)\\
&=\beta \sum_{[\boldsymbol{v}^I,\boldsymbol{v}^O]\in \{\boldsymbol{v}^{1}, ..., \boldsymbol{v}^{D}\}} \left( \mathbb{E}\left( \left.\frac{\partial E}{\partial \theta_j}\right| \boldsymbol{v^I},\boldsymbol{v^O} \right) - \mathbb{E}\left( \left.\frac{\partial E}{\partial \theta_j}\right|\boldsymbol{v^I}\right)\right) 
\end{aligned}}
\end{array}
\label{eq:gradient_funcapp_calc}
\end{equation}

\subsection{Hessian of KL Divergence}
Hessian of Log-likelihood for a single data is estimated first. It uses the fact that in the case of Ising type energy, $\frac{\partial^2 E}{\partial \theta_i \partial \theta_j}=0$ for all possible $i$ and $j$. 

\begin{equation}
\begin{array}{l}
{
\begin{aligned} \frac{\partial^2 \ln p}{\partial \theta_i \partial \theta_j} (\boldsymbol{v} )
=&-\beta \sum_{\boldsymbol{h}} \frac{\partial p(\boldsymbol{h} | \boldsymbol{v})}{\partial \theta_i} \frac{\partial E(\boldsymbol{v}, \boldsymbol{h})}{\partial \theta_j}+\beta \sum_{\boldsymbol{v}', \boldsymbol{h}'} \frac{\partial p(\boldsymbol{v}', \boldsymbol{h}')}{\partial \theta_i} \frac{\partial E(\boldsymbol{v}', \boldsymbol{h}')}{\partial \theta_j} \\
=&-\beta \sum_{\boldsymbol{h}} \frac{\partial}{\partial \theta_i} \left(\frac{ e^{-\beta E(\boldsymbol{v}, \boldsymbol{h})}}{\sum_{\boldsymbol{h}''} e^{-\beta E(\boldsymbol{v}, \boldsymbol{h}'')}}\right) \frac{\partial E(\boldsymbol{v}, \boldsymbol{h})}{\partial \theta_j}+\beta \sum_{\boldsymbol{v}', \boldsymbol{h}'} \frac{\partial }{\partial \theta_i}  \left(\frac{ e^{-\beta E(\boldsymbol{v}', \boldsymbol{h}')}}{\sum_{\boldsymbol{v}''',\boldsymbol{h}'''} e^{-\beta E(\boldsymbol{v}''', \boldsymbol{h}''')}}\right) \frac{\partial E(\boldsymbol{v}', \boldsymbol{h}')}{\partial \theta_j} \\
=&\beta^2 \sum_{\boldsymbol{h}} \left(p(\boldsymbol{h} | \boldsymbol{v})\frac{\partial E(\boldsymbol{v}, \boldsymbol{h})}{\partial \theta_i} - p(\boldsymbol{h} | \boldsymbol{v}) \sum_{\boldsymbol{h}''}  p(\boldsymbol{h}'' | \boldsymbol{v}) \frac{\partial E(\boldsymbol{v}, \boldsymbol{h}'')}{\partial \theta_i} \right) \frac{\partial E(\boldsymbol{v}, \boldsymbol{h})}{\partial \theta_j}\\
&-\beta^2 \sum_{\boldsymbol{v}', \boldsymbol{h}'}\left(p(\boldsymbol{v}',\boldsymbol{h}')\frac{\partial E(\boldsymbol{v}', \boldsymbol{h}')}{\partial \theta_i} - p(\boldsymbol{v}',\boldsymbol{h}') \sum_{\boldsymbol{v}''',\boldsymbol{h}'''}  p(\boldsymbol{v}''',\boldsymbol{h}''') \frac{\partial E(\boldsymbol{v}''', \boldsymbol{h}''')}{\partial \theta_i} \right) \frac{\partial E(\boldsymbol{v}', \boldsymbol{h}')}{\partial \theta_j} \\
=&\beta^2 \left(\sum_{\boldsymbol{h}} p(\boldsymbol{h} | \boldsymbol{v})\frac{\partial E(\boldsymbol{v}, \boldsymbol{h})}{\partial \theta_i}\frac{\partial E(\boldsymbol{v}, \boldsymbol{h})}{\partial \theta_j} \right)- \beta^2\left( \sum_{\boldsymbol{h}}  p(\boldsymbol{h} | \boldsymbol{v}) \frac{\partial E(\boldsymbol{v}, \boldsymbol{h})}{\partial \theta_i} \right)
\left( \sum_{\boldsymbol{h}}  p(\boldsymbol{h} | \boldsymbol{v}) \frac{\partial E(\boldsymbol{v}, \boldsymbol{h})}{\partial \theta_j} \right)\\
&-\beta^2 \left(\sum_{\boldsymbol{v}',\boldsymbol{h}'} p(\boldsymbol{v}',\boldsymbol{h}')\frac{\partial E(\boldsymbol{v}', \boldsymbol{h}')}{\partial \theta_i}\frac{\partial E(\boldsymbol{v}', \boldsymbol{h}')}{\partial \theta_j} \right)+ \beta^2\left( \sum_{\boldsymbol{v}', \boldsymbol{h}'}  p(\boldsymbol{v}', \boldsymbol{h}') \frac{\partial E(\boldsymbol{v}', \boldsymbol{h}')}{\partial \theta_i} \right)
\left( \sum_{\boldsymbol{v}', \boldsymbol{h}'}  p(\boldsymbol{v}', \boldsymbol{h}') \frac{\partial E(\boldsymbol{v}', \boldsymbol{h}')}{\partial \theta_j} \right)\\
=&\beta^2 \left(\operatorname{Cov}\left(\left.\frac{\partial E}{\partial \theta_i},\frac{\partial E}{\partial \theta_j}\right|\boldsymbol{v} \right)-\operatorname{Cov}\left(\frac{\partial E}{\partial \theta_i},\frac{\partial E}{\partial \theta_j} \right) \right)
\end{aligned}}
\end{array}
\end{equation}
The hessian of KL Divergence is estimated as: 
\begin{equation}
\begin{array}{l}
{
\begin{aligned} 
    \frac{\partial^2 D_{KL}(q||p)}{\partial \theta_i \partial \theta_j} &= \beta^2 \left(\operatorname{Cov}\left(\frac{\partial E}{\partial \theta_i},\frac{\partial E}{\partial \theta_j} \right)-\sum_{\boldsymbol{v}\in \{\boldsymbol{v}^{1}, ..., \boldsymbol{v}^{D}\}} q(\boldsymbol{v}) \operatorname{Cov}\left(\left.\frac{\partial E}{\partial \theta_i},\frac{\partial E}{\partial \theta_j}\right|\boldsymbol{v} \right) \right)
\end{aligned}}
\end{array}
\end{equation}

\subsection{Hessian of Negative Conditional Log-likelihood for a single data}
Hessian for a single data is estimated as: $\boldsymbol{v}\equiv [\boldsymbol{v}^I,\boldsymbol{v}^O]$

\begingroup
\allowdisplaybreaks
\begin{equation}
\begin{array}{l}
{
\begin{aligned} \frac{\partial^2 \mathcal{N}}{\partial \theta_i \partial \theta_j}
=&\beta  \left(
\sum_{\boldsymbol{h}''} \frac{\partial p(\boldsymbol{h}''|\boldsymbol{v}^I,\boldsymbol{v}^O)}{\partial\theta_i} \frac{\partial E(\boldsymbol{v}^I,\boldsymbol{v}^O, \boldsymbol{h}'')}{\partial \theta_j}-
\sum_{\boldsymbol{v'}^O,\boldsymbol{h}'} 
\frac{\partial p(\boldsymbol{v'}^O ,\boldsymbol{h}' | \boldsymbol{v}^I)}{\partial \theta_i} \frac{\partial E(\boldsymbol{v}^I,\boldsymbol{v'}^O, \boldsymbol{h}')}{\partial \theta_j} 
\right)\\
=&\beta  \left(
\sum_{\boldsymbol{h}''} \frac{\partial}{\partial\theta_i} 
 \left(\frac{ e^{-\beta E(\boldsymbol{v}^I,\boldsymbol{v}^O, \boldsymbol{h}'')}}{\sum_{\overline{\boldsymbol{h}}} e^{-\beta E(\boldsymbol{v}^I,\boldsymbol{v}^O, \overline{\boldsymbol{h}})}}\right)
\frac{\partial
E(\boldsymbol{v}^I,\boldsymbol{v}^O, \boldsymbol{h}'')}{\partial \theta_j} 
-\sum_{\boldsymbol{v'}^O,\boldsymbol{h}'} 
\frac{\partial}{\partial \theta_i}  
\left(\frac{ e^{-\beta E(\boldsymbol{v}^I,\boldsymbol{v'}^O, \boldsymbol{h}')}}{\sum_{\widetilde{\boldsymbol{v}}^O,\widetilde{\boldsymbol{h}}} e^{-\beta E(\boldsymbol{v}^I,\widetilde{\boldsymbol{v}}^O,\widetilde{ \boldsymbol{h})}}}\right)
\frac{\partial E(\boldsymbol{v}^I,\boldsymbol{v'}^O, \boldsymbol{h}')}{\partial \theta_j} 
\right)\\

=&-\beta^2 \left(
\sum_{\boldsymbol{h}''}
 \left(
 p(\boldsymbol{h}''|\boldsymbol{v}^I,\boldsymbol{v}^O)\frac{\partial E(\boldsymbol{v}^I,\boldsymbol{v}^O, \boldsymbol{h}'')}{\partial\theta_i} 
 -\sum_{\overline{\boldsymbol{h}}}  p(\overline{\boldsymbol{h}}|\boldsymbol{v}^I,\boldsymbol{v}^O)
 \frac{\partial E(\boldsymbol{v}^I,\boldsymbol{v}^O, \overline{\boldsymbol{h}})}{\partial\theta_i} 
 \right)
\frac{\partial
E(\boldsymbol{v}^I,\boldsymbol{v}^O, \boldsymbol{h}'')}{\partial \theta_j} 
 \right. \\& \qquad \qquad - \left.  
\sum_{\boldsymbol{v'}^O,\boldsymbol{h}'} 
 \left(
 p(\boldsymbol{v'}^O,\boldsymbol{h}'|\boldsymbol{v}^I)\frac{\partial E(\boldsymbol{v}^I,\boldsymbol{v'}^O, \boldsymbol{h}')}{\partial\theta_i} 
 -\sum_{\widetilde{\boldsymbol{v}}^O\widetilde{\boldsymbol{h}}}  p(\widetilde{\boldsymbol{v}}^O,\widetilde{\boldsymbol{h}}|\boldsymbol{v}^I)
 \frac{\partial E(\boldsymbol{v}^I,\widetilde{\boldsymbol{v}}^O, \widetilde{\boldsymbol{h}})}{\partial\theta_i} 
 \right)
\frac{\partial E(\boldsymbol{v}^I,\boldsymbol{v'}^O, \boldsymbol{h}')}{\partial \theta_j} 
\right)\\

=&-\beta^2 \left(
 \left(
\sum_{\boldsymbol{h}''}
 p(\boldsymbol{h}''|\boldsymbol{v}^I,\boldsymbol{v}^O)\frac{\partial E(\boldsymbol{v}^I,\boldsymbol{v}^O, \boldsymbol{h}'')}{\partial\theta_i} 
 \frac{\partial E(\boldsymbol{v}^I,\boldsymbol{v}^O, \boldsymbol{h}'')}{\partial \theta_j}\right)
 \right. \\& \qquad  \qquad
 -\left(\sum_{\overline{\boldsymbol{h}}}  p(\overline{\boldsymbol{h}}|\boldsymbol{v}^I,\boldsymbol{v}^O)
 \frac{\partial E(\boldsymbol{v}^I,\boldsymbol{v}^O, \overline{\boldsymbol{h}})}{\partial\theta_i} 
 \right)
 \left(\sum_{\boldsymbol{h}''}
 p(\boldsymbol{h}''|\boldsymbol{v}^I,\boldsymbol{v}^O)
\frac{\partial
E(\boldsymbol{v}^I,\boldsymbol{v}^O, \boldsymbol{h}'')}{\partial \theta_j} 
 \right)
  \\& \qquad  \qquad -
 \left(\sum_{\boldsymbol{v'}^O,\boldsymbol{h}'} 
 p(\boldsymbol{v'}^O,\boldsymbol{h}'|\boldsymbol{v}^I)\frac{\partial E(\boldsymbol{v}^I,\boldsymbol{v'}^O, \boldsymbol{h}')}{\partial\theta_i} 
 \frac{\partial E(\boldsymbol{v}^I,\boldsymbol{v'}^O, \boldsymbol{h}')}{\partial \theta_j} \right)
 \\& \qquad  \qquad  \left.  
 + \left(\sum_{\widetilde{\boldsymbol{v}}^O\widetilde{\boldsymbol{h}}}  p(\widetilde{\boldsymbol{v}}^O,\widetilde{\boldsymbol{h}}|\boldsymbol{v}^I)
 \frac{\partial E(\boldsymbol{v}^I,\widetilde{\boldsymbol{v}}^O, \widetilde{\boldsymbol{h}})}{\partial\theta_i} 
 \right)
  \left(\sum_{\boldsymbol{v'}^O,\boldsymbol{h}'} 
 p(\boldsymbol{v'}^O,\boldsymbol{h}'|\boldsymbol{v}^I)
\frac{\partial E(\boldsymbol{v}^I,\boldsymbol{v'}^O, \boldsymbol{h}')}{\partial \theta_j} 
\right)\right)\\

&=\beta^2 \left(\operatorname{Cov}\left(\left.\frac{\partial E}{\partial \theta_i},\frac{\partial E}{\partial \theta_j}\right|\boldsymbol{v}^I \right)-\operatorname{Cov}\left(\left.\frac{\partial E}{\partial \theta_i},\frac{\partial E}{\partial \theta_j}\right|\boldsymbol{v} \right) \right)
\end{aligned}}
\end{array}
\end{equation}
\endgroup

Now considering the visible data, $[\boldsymbol{v}^I,\boldsymbol{v}^O]\in \{\boldsymbol{v}^{1}, ..., \boldsymbol{v}^{D}\}$, 
\begin{equation}
\begin{array}{l}
{\begin{aligned}
\frac{\partial^2 \mathcal{N}}{\partial \theta_i \partial \theta_j}
=&\beta^2  \sum_{[\boldsymbol{v}^I,\boldsymbol{v}^O]\in \{\boldsymbol{v}^{1}, ..., \boldsymbol{v}^{D}\}} \left(\operatorname{Cov}\left(\left.\frac{\partial E}{\partial \theta_i},\frac{\partial E}{\partial \theta_j}\right|\boldsymbol{v}^I \right)-\operatorname{Cov}\left(\left.\frac{\partial E}{\partial \theta_i},\frac{\partial E}{\partial \theta_j}\right|\boldsymbol{v}^I,\boldsymbol{v}^O \right) \right)
\end{aligned}}
\end{array}
\end{equation}

\subsection{Derivative of KL Divergence w.r.t. Inverse temperature}
\begin{align*}
    \frac{d D_{KL}}{d \beta} & = -\sum_{\boldsymbol{v}\in \{\boldsymbol{v}^{1}, ..., \boldsymbol{v}^{D}\}} q(\boldsymbol{v})\frac{d}{d\beta}  \ln{\frac{p(\boldsymbol{v})}{q(\boldsymbol{v})}} = -\sum_{\boldsymbol{v}\in \{\boldsymbol{v}^{1}, ..., \boldsymbol{v}^{D}\}} \frac{q(\boldsymbol{v})}{p(\boldsymbol{v}) }
    \frac{d}{d\beta} \frac{\sum_{\boldsymbol{h}} e^{-\beta E(\boldsymbol{v},\boldsymbol{h})}}{\sum_{\boldsymbol{v}',\boldsymbol{h}'} e^{-\beta E(\boldsymbol{v}',\boldsymbol{h}')}} \\
    & = -\sum_{\boldsymbol{v}\in \{\boldsymbol{v}^{1}, ..., \boldsymbol{v}^{D}\}} \frac{q(\boldsymbol{v})}{p(\boldsymbol{v}) } \left(
    \frac{- \sum_{\boldsymbol{h}}  E(\boldsymbol{v},\boldsymbol{h}) e^{-\beta E(\boldsymbol{v},\boldsymbol{h})}}{\sum_{\boldsymbol{v}',\boldsymbol{h}'} e^{-\beta E(\boldsymbol{v}',\boldsymbol{h}')}}
    + \frac{\left(\sum_{\boldsymbol{h}} e^{-\beta E(\boldsymbol{v},\boldsymbol{h})}\right) \left(\sum_{\boldsymbol{v}',\boldsymbol{h}'} E(\boldsymbol{v}',\boldsymbol{h}') e^{-\beta E(\boldsymbol{v}',\boldsymbol{h}')}\right)
    }{\left(\sum_{\boldsymbol{v}'',\boldsymbol{h}''} e^{-\beta E(\boldsymbol{v}'',\boldsymbol{h}'')}\right)^2}
    \right)\\
    & = \sum_{\boldsymbol{v}\in \{\boldsymbol{v}^{1}, ..., \boldsymbol{v}^{D}\}} \frac{q(\boldsymbol{v})}{p(\boldsymbol{v}) } \left(
     \sum_{\boldsymbol{h}}  E(\boldsymbol{v},\boldsymbol{h})p(\boldsymbol{v},\boldsymbol{h}) 
    - p(\boldsymbol{v}) \sum_{\boldsymbol{v}',\boldsymbol{h}'} E(\boldsymbol{v}',\boldsymbol{h}') p(\boldsymbol{v}',\boldsymbol{h}')
    \right)\\
    & = \sum_{\boldsymbol{v}\in \{\boldsymbol{v}^{1}, ..., \boldsymbol{v}^{D}\}} q(\boldsymbol{v}) \left(
     \sum_{\boldsymbol{h}}  E(\boldsymbol{v},\boldsymbol{h})p(\boldsymbol{h}|\boldsymbol{v}) 
    - \sum_{\boldsymbol{v}',\boldsymbol{h}'} E(\boldsymbol{v}',\boldsymbol{h}') p(\boldsymbol{v}',\boldsymbol{h}')
    \right)\\
    & = -\mathbb{E}_{\boldsymbol{v},\boldsymbol{h}}(E)+ \sum_{\boldsymbol{v}\in \{\boldsymbol{v}^{1}, ..., \boldsymbol{v}^{D}\}} q(\boldsymbol{v}) 
     \sum_{\boldsymbol{h}}  E(\boldsymbol{v},\boldsymbol{h})p(\boldsymbol{h}|\boldsymbol{v})
\end{align*}
    
\begin{align*}
     \frac{d^2 D_{KL}}{d \beta^2} & =
     \sum_{\boldsymbol{v}\in \{\boldsymbol{v}^{1}, ..., \boldsymbol{v}^{D}\}} q(\boldsymbol{v}) \left(
     \sum_{\boldsymbol{h}}  E(\boldsymbol{v},\boldsymbol{h})\frac{d}{d\beta}p(\boldsymbol{h}|\boldsymbol{v}) 
    - \sum_{\boldsymbol{v}',\boldsymbol{h}'} E(\boldsymbol{v}',\boldsymbol{h}') \frac{d}{d\beta}p(\boldsymbol{v}',\boldsymbol{h}')
    \right)\\
    & =
     \sum_{\boldsymbol{v}\in \{\boldsymbol{v}^{1}, ..., \boldsymbol{v}^{D}\}} q(\boldsymbol{v}) \left(
     \sum_{\boldsymbol{h}}  E(\boldsymbol{v},\boldsymbol{h})\underbrace{\frac{d}{d\beta}
     \frac{ e^{-\beta E(\boldsymbol{v},\boldsymbol{h})}}{\sum_{\boldsymbol{h}''} e^{-\beta E(\boldsymbol{v},\boldsymbol{h}'')}} }_{\text{Term }I}
    - \sum_{\boldsymbol{v}',\boldsymbol{h}'} E(\boldsymbol{v}',\boldsymbol{h}') \underbrace{\frac{d}{d\beta}
    \frac{ e^{-\beta E(\boldsymbol{v}',\boldsymbol{h}')}}{\sum_{\boldsymbol{v}'',\boldsymbol{h}''} e^{-\beta E(\boldsymbol{v}'',\boldsymbol{h}'')}}}_{\text{Term }II }
    \right)
\end{align*}
Term $I$ is evaluated as: 
\begin{align*}
    \frac{d}{d\beta}
     \frac{ e^{-\beta E(\boldsymbol{v},\boldsymbol{h})}}{\sum_{\boldsymbol{h}''} e^{-\beta E(\boldsymbol{v},\boldsymbol{h}'')}} &=- 
     \frac{ E(\boldsymbol{v},\boldsymbol{h}) e^{-\beta E(\boldsymbol{v},\boldsymbol{h})}}{\sum_{\boldsymbol{h}''} e^{-\beta E(\boldsymbol{v},\boldsymbol{h}'')}}
     +
     \frac{ e^{-\beta E(\boldsymbol{v},\boldsymbol{h})}\sum_{\boldsymbol{h}'} E(\boldsymbol{v},\boldsymbol{h}')e^{-\beta E(\boldsymbol{v},\boldsymbol{h}')}}{\left(\sum_{\boldsymbol{h}''} e^{-\beta E(\boldsymbol{v},\boldsymbol{h}'')}\right)^2}\\
     &=-E(\boldsymbol{v},\boldsymbol{h})p(\boldsymbol{h}|\boldsymbol{v})+ p(\boldsymbol{h}|\boldsymbol{v})\sum_{\boldsymbol{h}'} E(\boldsymbol{v},\boldsymbol{h}')p(\boldsymbol{h'}|\boldsymbol{v})
\end{align*}
Term $II$ is evaluated as:
\begin{align*}
    \frac{d}{d\beta}
    \frac{ e^{-\beta E(\boldsymbol{v}',\boldsymbol{h}')}}{\sum_{\boldsymbol{v}'',\boldsymbol{h}''} e^{-\beta E(\boldsymbol{v}'',\boldsymbol{h}'')}} &=- 
     \frac{ E(\boldsymbol{v}',\boldsymbol{h}') e^{-\beta E(\boldsymbol{v}',\boldsymbol{h}')}}{Z}
     +
     \frac{ e^{-\beta E(\boldsymbol{v}',\boldsymbol{h}')}\sum_{\boldsymbol{v}'',\boldsymbol{h}''} E(\boldsymbol{v}'',\boldsymbol{h}'')e^{-\beta E(\boldsymbol{v}'',\boldsymbol{h}'')}}{Z^2}\\
     &=-E(\boldsymbol{v}',\boldsymbol{h}')p(\boldsymbol{v}',\boldsymbol{h}')+ p(\boldsymbol{v}',\boldsymbol{h}')\sum_{\boldsymbol{v}'',\boldsymbol{h}''} E(\boldsymbol{v''},\boldsymbol{h}'')p(\boldsymbol{v}'',\boldsymbol{h}'')
\end{align*}
Combining the two terms: 
\begin{align*}
    \frac{d^2 D_{KL}}{d \beta^2} & =  \sum_{\boldsymbol{v}\in \{\boldsymbol{v}^{1}, ..., \boldsymbol{v}^{D}\}} q(\boldsymbol{v}) \left(  \sum_{\boldsymbol{h}}  -E^2(\boldsymbol{v},\boldsymbol{h})p(\boldsymbol{h}|\boldsymbol{v})+ \left(\sum_{\boldsymbol{h}'} E(\boldsymbol{v},\boldsymbol{h}')p(\boldsymbol{h'}|\boldsymbol{v})\right)^2 \right.\\&\qquad \left.
    +\sum_{\boldsymbol{v}',\boldsymbol{h}'} E^2(\boldsymbol{v}',\boldsymbol{h}')p(\boldsymbol{v}',\boldsymbol{h}')- \left(\sum_{\boldsymbol{v}'',\boldsymbol{h}''} E(\boldsymbol{v''},\boldsymbol{h}'')p(\boldsymbol{v}'',\boldsymbol{h}'')\right)^2 \right)\\
    &= \sum_{\boldsymbol{v}\in \{\boldsymbol{v}^{1}, ..., \boldsymbol{v}^{D}\}} q(\boldsymbol{v}) \left( -\operatorname{Var}(E|\boldsymbol{v}) + \operatorname{Var}(E) \right)
\end{align*}

\subsection{Derivative of Negative Like-Likelihood w.r.t. Inverse temperature}

The derivative of log-likelihood of conditional probability for a single data, $\boldsymbol{v}\equiv [\boldsymbol{v}^I,\boldsymbol{v}^O]$ is calculated first:

\begin{align*}
    \frac{d \ln{p(\boldsymbol{v}| \boldsymbol{v}^I)}}{d \beta} & = \frac{d}{d\beta} \left(\ln{\frac{p(\boldsymbol{v})}{p(\boldsymbol{v}^I)}} \right)=  \frac{d}{d\beta} \left( \ln{ \sum_{\boldsymbol{h}} e^{-\beta E (\boldsymbol{v},\boldsymbol{h})}} - \ln{ \sum_{\overline{\boldsymbol{v}}^O,\boldsymbol{h}} e^{-\beta E (\boldsymbol{v}^I, \overline{\boldsymbol{v}}^O, \boldsymbol{h})}} \right)\\
    & =  \sum_{\boldsymbol{h}}  - E(\boldsymbol{v},\boldsymbol{h}) 
    \frac{e^{-\beta E(\boldsymbol{v},\boldsymbol{h})}}{\sum_{\boldsymbol{h}'} e^{-\beta E(\boldsymbol{v},\boldsymbol{h}')}}
    +  \sum_{\boldsymbol{v}^{O'},\boldsymbol{h}'} E(\boldsymbol{v}^I, \boldsymbol{v}^{O'},\boldsymbol{h}') \frac{ e^{-\beta E(\boldsymbol{v}^I, \boldsymbol{v}^{O'},\boldsymbol{h}')}
    }{ \sum_{\boldsymbol{v}^{O''},\boldsymbol{h}''} e^{-\beta E(\boldsymbol{v}^I, \boldsymbol{v}^{O''},\boldsymbol{h}'')}}
    \\
    & =  \sum_{\boldsymbol{h}}  - E(\boldsymbol{v},\boldsymbol{h}) 
   p(\boldsymbol{v}, \boldsymbol{h}|\boldsymbol{v})
    +  \sum_{\boldsymbol{v}^{O'},\boldsymbol{h}'} E(\boldsymbol{v}^I, \boldsymbol{v}^{O'},\boldsymbol{h}') p(\boldsymbol{v}^I, \boldsymbol{v}^{O'}, \boldsymbol{h}|\boldsymbol{v}^I) \\
    & = - \mathbb{E}(E|\boldsymbol{v}) + \mathbb{E}(E|\boldsymbol{v}^I) 
\end{align*}

The second derivative is estimated as
\begin{align*}
     \frac{d^2 \ln{p(\boldsymbol{v}| \boldsymbol{v}^I)}}{d \beta^2} & =
     \sum_{\boldsymbol{h}}  E(\boldsymbol{v},\boldsymbol{h})\underbrace{\frac{d}{d\beta}
     \frac{e^{-\beta E(\boldsymbol{v},\boldsymbol{h})}}{\sum_{\boldsymbol{h}'} e^{-\beta E(\boldsymbol{v},\boldsymbol{h}')}}}_{\text{Term }I}
    - \sum_{\boldsymbol{v}^{O'},\boldsymbol{h}'} E(\boldsymbol{v}^I, \boldsymbol{v}^{O'},\boldsymbol{h}') \underbrace{\frac{d}{d\beta}
     \frac{ e^{-\beta E(\boldsymbol{v}^I, \boldsymbol{v}^{O'},\boldsymbol{h}')}
    }{ \sum_{\boldsymbol{v}^{O''},\boldsymbol{h}''} e^{-\beta E(\boldsymbol{v}^I, \boldsymbol{v}^{O''},\boldsymbol{h}'')}}}_{\text{Term }II }
\end{align*}
Term $I$ is evaluated as: 
\begin{align*}
\frac{d}{d\beta}
     \frac{e^{-\beta E(\boldsymbol{v},\boldsymbol{h})}}{\sum_{\boldsymbol{h}'} e^{-\beta E(\boldsymbol{v},\boldsymbol{h}')}} &=- 
     \frac{ E(\boldsymbol{v},\boldsymbol{h}) e^{-\beta E(\boldsymbol{v},\boldsymbol{h})}}{\sum_{\boldsymbol{h}'} e^{-\beta E(\boldsymbol{v},\boldsymbol{h}')}}
     +
     \frac{ e^{-\beta E(\boldsymbol{v},\boldsymbol{h})}\sum_{\boldsymbol{h}'} E(\boldsymbol{v},\boldsymbol{h}')e^{-\beta E(\boldsymbol{v},\boldsymbol{h}')}}{\left(\sum_{\boldsymbol{h}''} e^{-\beta E(\boldsymbol{v},\boldsymbol{h}'')}\right)^2}\\
     &=-E(\boldsymbol{v},\boldsymbol{h})p(\boldsymbol{h}|\boldsymbol{v})+ p(\boldsymbol{h}|\boldsymbol{v})\sum_{\boldsymbol{h}'} E(\boldsymbol{v},\boldsymbol{h}')p(\boldsymbol{h'}|\boldsymbol{v})\\
     &=-E(\boldsymbol{v},\boldsymbol{h})p(\boldsymbol{h}|\boldsymbol{v})+ p(\boldsymbol{h}|\boldsymbol{v})\mathbb{E}(E|\boldsymbol{v})
\end{align*}
Term $II$ is evaluated as:
\begin{align*}
    \frac{d}{d\beta}
     \frac{ e^{-\beta E(\boldsymbol{v}^I, \boldsymbol{v}^{O'},\boldsymbol{h}')}
    }{ \sum_{\boldsymbol{v}^{O''},\boldsymbol{h}''} e^{-\beta E(\boldsymbol{v}^I, \boldsymbol{v}^{O''},\boldsymbol{h}'')}}  &=- 
     E(\boldsymbol{v}^I, \boldsymbol{v}^{O'},\boldsymbol{h}') \frac{  e^{-\beta E(\boldsymbol{v}^I, \boldsymbol{v}^{O'},\boldsymbol{h}')}}{ \sum_{\boldsymbol{v}^{O''},\boldsymbol{h}''} e^{-\beta E(\boldsymbol{v}^I, \boldsymbol{v}^{O''},\boldsymbol{h}'')} }
          + \\
          & \qquad 
    \frac{ e^{-\beta E(\boldsymbol{v}^I, \boldsymbol{v}^{O'},\boldsymbol{h}')}  \sum_{\boldsymbol{v}^{O''},\boldsymbol{h}''} E(\boldsymbol{v}^I, \boldsymbol{v}^{O''},\boldsymbol{h}'') e^{-\beta E(\boldsymbol{v}^I, \boldsymbol{v}^{O''},\boldsymbol{h}'') }}{\left( \sum_{\boldsymbol{v}^{O'''},\boldsymbol{h}'''} e^{-\beta E(\boldsymbol{v}^I, \boldsymbol{v}^{O'''},\boldsymbol{h}''')} \right)^2}      \\
     &=-E(\boldsymbol{v}^I, \boldsymbol{v}^{O'},\boldsymbol{h}') p(\boldsymbol{v}^{O'},\boldsymbol{h}' | \boldsymbol{v}^I ) + 
    p ( \boldsymbol{v}^{O'},\boldsymbol{h}' | \boldsymbol{v}^I )  \mathbb{ E } (E|\boldsymbol{v}^I) 
\end{align*}
Combining the two terms: 
\begin{align*}
    \frac{d^2 \ln{p(\boldsymbol{v}| \boldsymbol{v}^I)}}{d \beta^2} & =
     \sum_{\boldsymbol{h}} -E^2(\boldsymbol{v},\boldsymbol{h})p(\boldsymbol{h}|\boldsymbol{v})+ \mathbb{E}(E|\boldsymbol{v}) \sum_{\boldsymbol{h}} E(\boldsymbol{v},\boldsymbol{h}) p(\boldsymbol{h}|\boldsymbol{v})
    \\ &\quad  + \sum_{\boldsymbol{v}^{O'},\boldsymbol{h}'}  E^2(\boldsymbol{v}^I, \boldsymbol{v}^{O'},\boldsymbol{h}') p(\boldsymbol{v}^{O'},\boldsymbol{h}' | \boldsymbol{v}^I ) -
      \mathbb{ E } (E|\boldsymbol{v}^I)  \sum_{\boldsymbol{v}^{O'},\boldsymbol{h}'} E(\boldsymbol{v}^I, \boldsymbol{v}^{O'},\boldsymbol{h}')p ( \boldsymbol{v}^{O'},\boldsymbol{h}' | \boldsymbol{v}^I ) \\
      & = - \mathbb{E}(E^2|\boldsymbol{v}) + \mathbb{E}^2(E|\boldsymbol{v}) + \mathbb{E}(E^2|\boldsymbol{v}^I) - \mathbb{E}^2(E|\boldsymbol{v}^I) \\
      & = -\operatorname{Var}(E|\boldsymbol{v}) + \operatorname{Var}(E|\boldsymbol{v}^I) 
\end{align*}

The first derivative of Negative conditional log-likelihood is estimated as: 
\begin{align*}
    \frac{d \mathcal{N}}{d \beta} & = \sum_{[\boldsymbol{v}^I,\boldsymbol{v}^O]\in \{\boldsymbol{v}^{1}, ..., \boldsymbol{v}^{D}\}} \mathbb{E}(E|\boldsymbol{v}) - \mathbb{E}(E|\boldsymbol{v}^I) 
\end{align*}

The second derivative of Negative conditional log-likelihood is estimated as: 
\begin{align*}
    \frac{d^2 \mathcal{N}}{d \beta^2} & = \sum_{[\boldsymbol{v}^I,\boldsymbol{v}^O]\in \{\boldsymbol{v}^{1}, ..., \boldsymbol{v}^{D}\}} -\operatorname{Var}(E|\boldsymbol{v}) + \operatorname{Var}(E|\boldsymbol{v}^I) 
\end{align*}

\section{Change of basis}\label{sec:basis}
It is a common practice to define the Ising states as either $\{0,1\}$ or $\{-1,+1\}$ states. The latter format is employed on the DWave machine. Here, the details about conversion between these formats are presented. For the purpose of discussion, the $\{0,1\}$ Ising model is represented with variables, $\{\boldsymbol{S},
\{H_i\}_{i=1}^{N_V},\{J_i\}_{i=1}^{N_C}\}$ and the $\{-1,1\}$ Ising with model is represented with overlined variables, $\{\overline{\boldsymbol{S}},
\{\overline{H}_i\}_{i=1}^{N_V},\{\overline{J}_i\}_{i=1}^{N_C}\}$. The three variables represent the state, field energy and interaction energy respectively. The states of the system can be interchanged using the following equation:
\begin{gather}
    \overline{\boldsymbol{S}} = 2\boldsymbol{S}-1\label{eq:state_transform}
\end{gather}
The interaction parameters can be interchanged as: 
\begin{gather}
    \overline{J}_k = \frac{1}{4}J_k\label{eq:interaction_transform}
\end{gather}
And the field parameter as:
\begin{gather}
    \overline{H}_{i} = \frac{1}{2} H_i + \frac{1}{4}\sum_{\pi(k,1)=i \text{ OR } \pi(k,2)=i} J_k \label{eq:parameter_transform}
\end{gather}
This transformation shifts the energy of each state with a constant value and hence leaves the Boltzmann probability unchanged as required. 

\end{document}